\documentclass[11pt,a4paper]{article}
\pdfoutput=1
\usepackage{jheppub}
\usepackage{appendix}
\usepackage{amssymb,amsmath,amstext,amsfonts,empheq}
\usepackage{bbold}
\usepackage{hyperref}
\counterwithout{equation}{section} 
\usepackage{simpler-wick}

\usepackage{bm}
\usepackage{graphicx}
\usepackage{xcolor}

\usepackage{overpic}
\usepackage{subcaption}
\usepackage{braket}

\usepackage[extra]{tipa}
\renewcommand{\arraystretch}{2}
\newcommand{\be}{\begin{equation}}
\newcommand{\ee}{\end{equation}}
\newcommand{\bea}{\begin{eqnarray}}
\newcommand{\eea}{\end{eqnarray}}

\newcommand{\bal}{\begin{aligned}}
\newcommand{\eal}{\end{aligned}}
\newcommand{\Lag}{{\mathcal{L}}}
\newcommand{\Lt}{{\mathcal{L}^{(3)}}}
\newcommand{\pp}{{P}}
\newcommand{\lam}{\lambda}
\newcommand{\acof}{\mathcal{A}}
\newcommand{\Hamt}{{\mathcal{H}^{(3)}}}
\newcommand{\Ham}{{\mathcal{H}}}

\newcommand{\gamm}{{\gamma}}
\newcommand{\Mp}{M_{\rm Pl}}
\newcommand{\llangle}{\langle \mkern-4mu \langle}
\newcommand{\rrangle}{\rangle \mkern-4mu\rangle}

\newcommand{\llangleb}{\Big\langle \mkern-8mu \Big\langle}
\newcommand{\rrangleb}{\Big\rangle \mkern-8mu\Big\rangle}

\newcommand{\Ima}{\mathrm{Im}}
\newcommand{\Rea}{\mathrm{Re}}

\newcommand{\Ha}{H}

\newcommand{\Lagp}{\mathcal{L}_{\partial'}}

\newcommand{\R}{\zeta}

\newcommand{\Rh}{\hat{\zeta}}

\newcommand{\bk}{\boldsymbol{k}}
\newcommand{\bK}{\boldsymbol{K}}
\newcommand{\bp}{\boldsymbol{p}}

\definecolor{alizarin}{rgb}{0.82, 0.1, 0.26}

\newcommand{\laa}{\Big\langle}
\newcommand{\raa}{\Big\rangle}

\newcommand{\p}{\mathcal{P}}

\newcommand{\Wr}{\mathcal{W}_{\tau_1}}
\usepackage{soul}
\definecolor{lgray}{gray}{0.90}
 
\renewcommand{\arraystretch}{2}

\newcommand{\bx}{\boldsymbol{x}}
\definecolor{alizarin}{rgb}{0.82, 0.1, 0.26}

\newcommand{\tauf}{\tau_{\rm 0}}

\title{ \centering \boldmath
Absence of one-loop effects on large scales from small scales in non-slow-roll dynamics II:\\ [4mm]

\Large{Quartic interactions and consistency relations}}
\author{\Large Jacopo Fumagalli\\} 
\affiliation{Departament de F\'isica Quàntica i Astrofísica and 
Institut de Ciències del Cosmos (ICC), \\[1mm]
Universitat de Barcelona,\\[1mm] Martí i Franquès 1, 08028 Barcelona, Spain}
\emailAdd{jfumagalli@fqa.ub.edu}
\abstract{ We prove explicitly the absence of one-loop corrections to large scales from small scales in transient non-slow-roll dynamics. Specifically, we address loop corrections to the primordial power spectrum, relative to tree-level, that are independent of the ratio between the two scales. We review all the necessary components, adapted to our context, to express one-loop diagrams as three-point functions, emphasizing the crucial role played by quartic interactions. Notably, we include the quartic Hamiltonian induced by the cubic Lagrangian and quartic interactions that are ensured by diffeomorphism invariance.  We then explicitly prove consistency relations for an arbitrary transient non-slow-roll phase involving operators with (time) derivatives. Finally, we calculate one-loop corrections by including contributions from the relevant cubic and quartic interactions, and express the final result as a total derivative term over comoving momenta, utilizing the consistency relations we established. This leads us to conclude that one-loop corrections to long-wavelength modes are unaffected by the physics of short and enhanced modes in non-slow-roll dynamics.

}

\begin{document}
\hfill{\flushright {}}
\maketitle
\newpage
\section{Introduction}

The foundational role of inflation in modern cosmology hinges on the expectation that the comoving curvature perturbation $\R$ remains constant on scales much larger than the horizon \cite{Weinberg:2003sw,Wands:2000dp,Lyth:2004gb}. This means that regardless of what occurs to perturbations on scales near the Hubble radius, the (very)long-wavelength modes would be unaffected. This is how the theory remains predictive and retains its allure. 

Although several opportunities to explore different inflationary epochs are envisaged in the future--see for instance \cite{LISACosmologyWorkingGroup:2022jok,Braglia:2024kpo}, we currently have access to only a small segment of the inflationary history. This corresponds to times when the scales observed in the Cosmic Microwave Background (CMB) or in Large-Scale Structure surveys are assumed to cross the horizon during inflation. We do not know what might have occurred between that era and the end of inflation, nor the onset of reheating. 
It would therefore be preferable to have the option to remain agnostic regarding these parts of the dynamics.

This is particularly relevant in the context of primordial black holes \cite{Zeldovich:1967lct,Hawking:1971ei,Carr:1974nx,Meszaros:1974tb,Carr:1975qj,Ivanov:1994pa,Garcia-Bellido:1996mdl} and scalar-induced gravitational waves \cite{Acquaviva:2002ud,Ananda:2006af,Mollerach:2003nq} arising from inflation. In such scenarios, it is often necessary to assume a phase that drives the system out of the attractor phase, resulting in an amplification of primordial perturbations at small scales relative to larger scales, such as those probed by the CMB.
Primary examples include non-slow-roll phases during the inflationary dynamics, such as ultra-slow-roll (USR) phases \cite{Tsamis:2003px,Kinney:2005vj,Namjoo:2012aa,Garcia-Bellido:2017mdw,Germani:2017bcs,Motohashi:2017kbs,Ballesteros:2017fsr,Hertzberg:2017dkh} or constant-roll \cite{Motohashi:2014ppa, Inoue:2001zt,Tzirakis:2007bf}. 
Within these frameworks, the enhanced short modes can potentially source at nonlinear level, i.e. through loop diagrams, long-wavelength modes.

Considering loop corrections to primordial correlators where interactions are localized in time and the external IR momenta are much smaller than the higher momenta running in the loop, one finds what we label as \textit{relative scale-invariant corrections}. These are contributions, relative to the tree-level one, that are independent of the ratio between short and long scales.\footnote{Note that there are examples of ``infrared cascades",
where long-wavelength modes are enhanced by short modes \cite{Barnaby:2009mc,Pearce:2017bdc,Fumagalli:2023loc}.  These enhancements are localized around a specific physical scale, typically the one of the horizon when a certain feature becomes relevant. This is different from the case studied in the current paper. Here we focus on arbitrarily long-wavelength modes that can be far from any particular feature or scale in the problem.} 
In addition, since we examine diagrams where the external momentum is well outside the horizon at the time interactions are active, it also follows that these contributions would effectively induce---surprisingly---a time dependence of \(\R\) on super-Hubble scales.
Note that the so called time dependence of the power spectrum at loop level was noted in early works \cite{Weinberg:2005vy,Kahya:2010xh} and it is a general issue not specific to a non-slow-roll phase. When considering a subset of diagrams in that context, one observes a specific manifestation of this general fact.

Within an ultra-slow-roll phase, how loop corrections of short enhanced modes influence arbitrarily large scales has been recently scrutinized by several authors. Whether these corrections, when considered individually, can become large enough to undermine the predictivity of inflationary scenarios related to PBH has been studied, for instance, in \cite{Kristiano:2022maq,Riotto:2023hoz,Kristiano:2023scm,Riotto:2023gpm,Firouzjahi:2023aum,Firouzjahi:2023ahg,Franciolini:2023lgy,Tasinato:2023ukp,Cheng:2023ikq,Maity:2023qzw,Davies:2023hhn,Kristiano:2024vst}, while other works question the very existence of these corrections \cite{Fumagalli:2023hpa,Tada:2023rgp,Inomata:2024lud}.
Different frameworks have also been pursued and proposed to study loop corrections in this context. Notably, using the separate universe approach \cite{Firouzjahi:2023ahg, Iacconi:2023ggt}, working in a different (spatially flat) gauge \cite{Inomata:2024lud, Ballesteros:2024zdp}, or even more explorative approaches using canonical transformations in the full phase space to avoid the generation of total derivative terms \cite{Braglia:2024zsl}.

Complementing our previous work \cite{Fumagalli:2023hpa}, we continue to pursue a standard in-in approach in the so called comoving or $\zeta$ gauge. We find that sufficient for the purpose of studying the effect of short-wavelength modes on long-wavelengths. This is, for instance, the language in which the quest concerning loop corrections in models with enhanced scalar perturbations was originally formulated  in \cite{Kristiano:2022maq}. There a single operator proportional to the derivative of the second slow-roll parameter was considered, leading to what we denote as a relative scale-invariant correction. In \cite{Fumagalli:2023hpa}, we emphasize the importance of total derivative terms in computing loop corrections. Their role in loop computations involving non-slow-roll phases was later emphasized also in \cite{Tada:2023rgp}, and more broadly in \cite{Kawaguchi:2024lsw,Braglia:2024zsl}. Specifically, in \cite{Fumagalli:2023hpa}, we consider a particular total derivative operator whose contribution cancels the one originally pointed out in \cite{Kristiano:2022maq}. However, as correctly indicated in \cite{Firouzjahi:2023bkt}, that contribution alone is not sufficient to grant the entire cancellation of all relative scale-invariant corrections.\footnote{In particular, in \cite{Fumagalli:2023hpa} we neglected terms proportional to the Green's function associated with the linear modes. As discussed in Sec. \ref{sec:origin scale invariant}, these terms also lead to relative scale-invariant correction to the power spectrum.} That is the aim of the current manuscript.

Once the generality of the problem is recognized, we follow and extend results proven in the framework of single clock inflation in \cite{Pimentel:2012tw} and apply them to the issue at hand.  We thus prove explicitly the absence of relative scale-invariant loop corrections to the primordial power spectrum for generic non-slow-roll dynamics with the strategy outlined in the next section.

\section{Roadmap and results}

We aim to prove the absence of loop corrections from short scales to large scales inducing a super-horizon time-dependence to the power spectrum of the primordial curvature perturbation $\R$, in transient non-slow-roll dynamics. In particular, we examine correction of the form:
\begin{align}
\p_{\R}^{\mathrm{1-loop}}(p,\tau) = \p^{\mathrm{tree}}(p,\tau)\int^{\tau} d\tau_1 \int d k \, C(k,\tau_1) 
\end{align}
where $\p^{\mathrm{tree}}$ is the tree-level dimensionless power spectrum, we label with $(\bp,\bp' )$ and $(\bk,\bk')$ long and short modes respectively, i.e. $\bp \ll \bk$, and $C(k,\tau_1)$ is a dimensionless quantity that depends only on the short modes. 

Following \cite{Pimentel:2012tw}, we discuss under which conditions loop diagrams can be formally expressed as the integral over time of three-point functions involving two short modes and one long mode. To serve our needs, we then want to apply the consistency relations \cite{Maldacena:2002vr,Creminelli:2004yq,Cheung:2007sv} to these latter. 
In doing that, key is the role played by quartic interactions. In particular, the quartic Hamiltonian induced from the cubic Lagrangian is needed to build three-point functions with time-derivative operators inside the loops, and quartic interaction whose presence is guaranteed by a residual diffeomorphism invariance (see Sec. \ref{sec:diffinduced}) are crucial to guarantee that consistency relations (for operators with spatial derivatives) are satisfied.\footnote{We do not explicitly include one type of contributions in our analysis. These are the diagrams coming from terms in the quartic Lagrangian which are not implied by the residual diff. invariance mentioned in Sec. \ref{sec:diffinduced}. As it was shown generically in \cite{Pimentel:2012tw}, the one-loop contribution from these terms cancels exactly with the one coming from the terms non-linearly induced by the tadpole counterterms. These latter start linearly in the perturbations, as  mentioned in Sec. \ref{sec:tad}, but unavoidably lead to non-linear terms as a consequence of the non-linear realization of time diffeomorphisms.}
Once including these diagrams one can write the generic relative scale-invariant one-loop correction schematically as 

\be\label{main}
\p^{\mathrm{1-loop}}(p)\propto\int^{\tau}_{\tauf}   d\tau_1\int  \frac{d\bk }{(2\pi)^3} \frac{p^3}{2\pi^2} \llangleb \frac{\delta\Lt}{\delta \R^A}\Big|_{\bk,-\bk}\Rh_{\bp} \rrangleb \propto \p^{\mathrm{tree}}(p)\int^{\tau}_{\tauf}   \int d\ln k \frac{d}{d\ln k}
\llangleb \frac{\delta \Lt}{\delta \R^A}\Big|_{\bk}\rrangleb,
\ee
where $\Lt$ is the cubic Lagrangian density and $\frac{\delta \Lt}{\delta \R^A}$ the two-field operator obtained by taking its functional derivative with respect to one of the operators building $\Lt$. For instance, $\R^A \equiv \{ \R, \R',\partial \R \}$. We consider the effect of short modes running in the loops whose interactions are active after a given time $\tauf$.
The integrand on the left-hand side of the previous equation is then the three-point function of two short modes (corresponding to the momenta running in the loop)  and one long mode (corresponding to the external IR scale). That leads to the expression on the right-hand side under the assumption that Maldacena consistency relations hold at all times of integration and contributions from the quartic interaction implied by residual diffeomorphism invariance, here understood on the left-hand side, are included (see Sec. \ref{sec:diffinduced}).

The total derivative allows then to trivially perform integration over comoving momenta. As a result, the one-loop correction would depend only on two arbitrary boundary scales, \(k_{\rm IR}\) and \(k_{\rm UV}\), which encompass the short enhanced modes.

The philosophy under the previous statement is similar to the one advocated, in this context, in Ref. \cite{Tada:2023rgp}, but with two important differences. First, as mentioned earlier, to arrive at the previous expression the addition of quartic interactions is decisive. Second, the structure of the two-point function is determined by the structure of the cubic Lagrangian, that is also crucial in drawing our conclusions.
As we will also show explicitly, the full $\Lt $ does contain only operators with at least either two time or two spatial derivatives. Therefore, the two point function of $\frac{\delta \Lt}{\delta \R^A}$ has always at least one derivative operator. This ensures the suppression of the integrand evaluated at the lower extrema by taking $k_{\rm IR}$ sufficiently small, i.e. well outside the horizon at $\tauf$. 
We are thus left with the integrand evaluated at the arbitrary cutoff scale $k_{\rm UV}$.\footnote{In scenarios where modes are boosted by several order of magnitudes around a given scale, it is sometimes customary to evaluate the integral within a window encompassing only the enhanced scales. This is morally allowed if the integral is independent upon varying the two cutoffs (until one hits the deep-horizon modes that should always be renormalized away). This is clearly not true if the integral over momenta is given by a total derivative.} 
This is the contribution from deep sub-Hubble modes at times interactions are considered. As advocated, for instance in \cite{Pimentel:2012tw,Tada:2023rgp}, these contributions can be set to zero by standard dimensional regularization. We do not elaborate further on this point. That would be particularly interesting when considering loop corrections from short to short scales. We emphasize that, once all one-loop contributions to the long-wavelength modes $\bp$ can be recasted in the form \eqref{main}, it is then clear that there is no dependence on the physics of the short modes affected by the non-slow-roll phase, i.e. the one-loop corrections to the long-wavelength are insensitive to the physics of the enhanced modes.  

We thus focus our efforts in proving explicitly Eq. \eqref{main} in the context of non-slow-roll dynamics.
Below, we provide a summary of the steps we pursued for our scope, highlighting the new results derived along the way:
\begin{itemize}
\item 
In Sec. \ref{sec:origin scale invariant}, we first set the stage by defining properly what we mean by relative scale-invariant corrections and show how they do arise in general.
\item In Sec. \ref{sec:oneloopthree}, we review all the required ingredients, adapted to our needs, to express one-loop diagrams as three-point functions. We find necessary to add a technical extension to results in  \cite{Pimentel:2012tw}. 
From the quartic induced Hamiltonian, we note spurious contributions, which come in addition to the terms needed to make a particular type of connected three-point function. Interestingly, we were able to show that these corrections cancel exactly once we include diagrams from the quadratic Hamiltonian induced by tadpole counterterms---see Sec. \ref{sec:tad}.
\item 
In Sec. \ref{sec:Hamiltonian}, we single out the relevant terms of the cubic and quartic Hamiltonian relevant for computing corrections to the long-wavelength modes for a phase of constant $\eta$ (at first order in $\epsilon$). We discuss the role of total derivative terms and outline equivalent ways of expressing the cubic Hamiltonian, suited for different purposes.
Specifically, these include deriving the relevant part of the quartic Hamiltonian, computing three-point functions, and assessing one-loop corrections to the power spectrum. 
\item 
In Sec. \ref{sec:consistency}, we prove explicitly (and analytically) consistency relations for an arbitrary transient non-slow-phase for operators with (time) derivatives. These are the ones needed for our purpose. Our findings are valid for short modes which are not necessarily evaluated outside the horizon and long modes well outside the horizon during the non-slow-roll phase.\footnote{This has not to be confused, and it is not in contradiction, with the well known violation of Maldacena consistency relations for non-attractor backgrounds \cite{Namjoo:2012aa,Martin:2012pe,Mooij:2015yka}. Here, we are considering the soft mode well outside the horizon during the non-slow-roll phase so that its power spectrum can be considered constant therein, i.e. the effect of the would be growing mode is negligible.} To achieve that, we use standard in-in formalism to compute three-point functions from the previously highlighted cubic interactions. We then exploit properties of the Hankel functions present in the generic expressions for the mode functions to recast derivatives of the two point correlators. We comment on the parts that are dependent on the smoothness of the transition and the generic ones that only rely on a phase of constant $\eta$. 
These results expand upon the work of \cite{Motohashi:2023syh} where consistency relations for the squeezed bispectrum $\langle \Rh\Rh\Rh\rangle $ were proven in transient non-slow-roll dynamics.   
\item 
In Sec. \ref{sec:explicitoneloop}, we compute one-loop corrections to the long-wavelength power spectrum in non-slow dynamics using the Hamiltonian detailed in Sec. \ref{sec:Hamiltonian}. Building upon our previous work \cite{Fumagalli:2023hpa}, we include the contributions from total derivative terms and discuss how to pass from one set of interactions to other equivalent ones. We further include quartic interactions, and tadpole induced contributions.
This computation allows us to show explicitly how all diagrams sum up in total derivatives over comoving momenta which we will be able to recognize using the consistency relations proven in Sec. \ref{sec:consistency}. Specifically, how the final result can be reorganized as detailed in Eq. \eqref{main}.

\end{itemize}

\section*{Notations, definitions and conventions}
\addcontentsline{toc}{section}{Notations, definitions and conventions}

Our final goal is to compute the one-loop correction to the dimensionless primordial power spectrum of the comoving curvature perturbation $\R$ at the end of inflation $\tau$ and defined as
\be\label{defpower}
\langle \hat{\R}_{\bp}(\tau) \hat{\R}_{\bp'}(\tau) \rangle = \left( 2\pi \right)^3 \delta \left( \bp + \bp' \right)  \frac{2\pi^2}{k^3}\mathcal{P}_\R(p,\tau).
\ee
To compute equal-time correlators, we use the standard in-in formalism \cite{Schwinger:1960qe,Jordan:1986ug,Calzetta:1986ey}. In particular, we consider interactions that become relevant only after a preferred time \(\tauf\). Consequently, the lower limit of integration in the in-in formula does not introduce any subtleties associated with the \( -\infty_\pm = -\infty (1 \pm i \epsilon) \) prescription and the $n\text{-point correlator}$ can be rewritten in terms of a series of nested commutators as detailed below:
\begin{equation}\label{fullinin}
\langle\Rh^{n}(\tau)\rangle=\sum_{j=0}^{\infty}i^{j}\int_{\tauf}^{\tau}d\tau_{1}\int_{\tauf}^{\tau_{1}}d\tau_{2}..\int_{\tauf}^{\tau_{j-1}}d\tau_{j}\langle[\Ha_{I}(\tau_{j}),[\Ha_{I}(\tau_{j-1}),..[\Ha_{I}(\tau_{1}),\Rh_{I}^{n}(\tau)]..]\rangle,
\end{equation}
the subscript $I$ labels fields in the interaction picture, i.e. fields evolving with the linear equation of motions. $H_I$ is the interaction Hamiltonian. To keep the notation concise, we omit this subscript on the right-hand sides of the correlator's formula, reintroducing it only when necessary. Therefore, the label $\Rh_{\bp}$ on the left and right sides of in-in formulas, such as Eq. \eqref{fullinin}, will not be equivalent. Additionally, we often omit the explicit time dependence of the correlated fields, such that, for example, $\langle \hat{\R}_{\bp}(\tau) \hat{\R}_{\bp'}(\tau) \rangle$ is denoted simply as $\langle \hat{\R}_{\bp} \hat{\R}_{\bp'} \rangle$ and the time dependence of the dimensionless power spectrum when evaluated at late time.

We are interested in studying the impact of short scales on long-wavelengths. We thus always work under the approximation $p \ll k$, where we denote with $\bk$ the short modes running in the loop, and with $\bp$ the external momenta associated with long-wavelength modes.
As is customary, bold symbols (e.g. $\bp,\bk$) are used to represent three vectors, while their corresponding magnitudes are denoted by the same letters in unbolded form (e.g. $p,k$). We use the Fourier transform convention $
f(\bx)=(2\pi)^{-3}\int d\bk\, e^{-i\bk\cdot\bx}f(\bk)$. 

In momentum space, canonical quantization of the free fields leads to
\be\label{interactionfield}
\Rh_{\bp}(\tau) =\R_{p}(\tau) \hat{a}(\bp) + \R^*_{p}(\tau) \hat{a}^{\dagger}(-\bp),
\ee
with annihilation/creation operators satisfying standard commutation relations
\be\label{quantiz}
[\hat{a}(\bp),\hat{a}^{\dagger}(\bp')]= (2\pi)^3 \delta(\bp -\bp'),
\ee
and $\R_p(\tau)$ labelling the mode functions of the free fields, solutions of the linear equations of motion. 
$\p_\R$ is computed perturbatively, i.e.
\be
\p_\R = \p_\R^{\mathrm{tree}} + \p_\R^{\mathrm{1-loop}}+ ...\,,
\ee
where the tree level power spectrum $\p_\R^{\mathrm{tree}}$, obtained by taking the correlators of linear fields, is given by
\be\label{treelevel}
\mathcal{P}_\R^{\mathrm{tree}}= \frac{p^3}{2\pi^2}|\R_p|^2.
\ee
To highlight the commutation structure, when dealing with expansions of Eq. \eqref{fullinin}, we occasionally use the same index for operators that originate from the same vertex, i.e.
\be\label{compact}
\left[ \Rh_{\bk_{2,1}}\Rh_{\bk_{2,2}}\Rh_{\bk_{2,3}}\Big|_{\tau_2},  \left[ \Rh_{\bk_{1,1}}\Rh_{\bk_{1,2}}\Rh_{\bk_{1,3}}\Big|_{\tau_1}, \Rh_{\bp}\Rh_{\bp'}\Big|_{\tau} \right] \right]\equiv \left[ \Rh^ 3_{2},  \left[ \Rh_1^3, \Rh_{\bp}\R_{\bp'} \right] \right]. 
\ee
and define
\be\label{bigK}
\int d\bK \equiv \int d\bK_1\int d\bK_2 
\ee
with
\be\label{bigK2}
\int d\bK_1\equiv \prod_{i=1}^{3}\left[\int \frac{d\bk_{1,i}}{(2\pi)^ 3}\right](2\pi)^ 3\delta\left(\sum_{i=1}^{3} \bk_{1,i}\right),\quad \int d\bK_2 =  \prod_{i=1}^{3}\left[\int \frac{d\bk_{2,i}}{(2\pi)^ 3}\right] (2\pi)^ 3 \delta\left(\sum_{i=1}^{3} \bk_{2,i}\right).
\ee
We often use a prime notation on the commutators to indicate that the Dirac delta function and $(2\pi)^3$ factors have been stripped from the commutator between two operators, i.e.: 
\begin{align}
\left[ \Rh_{\bk}(  \tau_1) ,\,  \Rh'_{\bp}(  \tau) \right] =&
\left( 2 \pi \right)^3 \delta^{(3)} \left( \bk + \bp \right) \left[ \Rh_{\bk}(  \tau_1) ,\,  \Rh'_{\bp}(  \tau) \right]',
\label{commutatornotation}
\end{align}
with the same purpose we also use a double brackets on the correlators:
\be\label{defdoublebracket}
\langle \Rh_{\bp} \Rh_{\bk} \Rh_{\bk'} \rangle = (2\pi)^3\delta(\bp+\bk+\bk' )\llangle \Rh_{\bp} \Rh_{\bk} \Rh_{\bk'} \rrangle.
\ee
We set $\Mp = 1$ and use curly letters to denotes Hamiltonian and Lagrangian density, i.e. $S = \int d\tau d^3 x  \mathcal{L}$ and $H= \int d^3 x \mathcal{H}$.
Finally we use the following definition for the slow-roll parameters:
\be
\epsilon = -\frac{\dot{H}}{H^2},\qquad \eta = \frac{\dot{\epsilon}}{\epsilon H},
\ee
with the dot indicating differentiation with respect to physical time, $H=\dot{a}/a$ is the Hubble rate, and $a$ denotes the scale factor of a spatially flat FLRW metric, $ds^2 = a(\tau)^2 (-d\tau^2 + d\bx ^2)$. 

\section{Origin of relative scale-invariant one-loop corrections and time dependence}\label{sec:origin scale invariant}
As we show in a moment, the super-horizon limits of the commutator relations in Eqs.  \eqref{commutatorsapprox} given below, lead to relative scale-invariant corrections to the one-loop power spectrum whenever a subset of interactions is considered. Note what we mean by \textit{relative scale-invariant corrections}: these refer to scenarios where, when examining the effects of short modes on long-wavelengths, the contributions to the one-loop dimensionless power spectrum, relative to tree-level, are independent of the external momentum denoted by $p$. We consider diagrams where the external momentum \(\bp\) is well outside the horizon at the time the interactions are active, so effectively these contributions would induce a time dependence of \(\R\) on super-Hubble scales.

In this section, we illustrate how these corrections arise in general without committing to a specific non-slow-roll phase.

\subsection{Commutator relations}
The equal-time commutator between the field $\Rh$ and its first derivative is given by the Wronskian detailed below ($W[f,g] = fg' -f'g$.):
\begin{align}
\left[ \Rh_{\bk}(  \tau_1) ,\,  \Rh'_{\bp}(  \tau_1) \right]' =&W[\R_p(\tau_1),\R_p^*(\tau_1)]\equiv
\Wr 
\label{commutator01}
\end{align}
where
\be\label{Wronskian}
\Wr  
=2i\, \Ima(\R_p(\tau_1)\R'^*_p(\tau_1)) = \frac{i}{2    \epsilon (\tau_1) \,a^2(\tau_1)},
\ee
and prime over commutators indicates the omission of the Dirac delta function and $(2\pi)^3$ factors, as defined in Eq. \eqref{commutatornotation}, while the prime over $\R$ denotes differentiation with respect to the conformal time $\tau$. Unequal-time commutators can also be expressed generally and concisely in terms of the Wronskian and the Green's function associated with the linear equations of motion as:\footnote{Specifically, the Green's function corresponds to Eq. \eqref{green} multiplied by \(\theta(\tau-\tau_1)\). Given the nested integrals that arise from time ordering, where \(\tau\) is always larger than \(\tau_1\), commutators like the one mentioned above will effectively represent the Green's function.}
\begin{align}\label{commutatorsgreen}
\left[ \Rh_{\bk}(  \tau_1) ,\,  \Rh_{\bp}(  \tau) \right]' =
\Wr \, g_p \left( \tau ,\, \tau_1 \right) ,\qquad \left[ \Rh'_{\bk}(  \tau_1) ,\,  \Rh_{\bp}(  \tau) \right]' =
\partial_{\tau_1}
\left( \Wr g_p \left( \tau ,\, \tau_1 \right) \right)
\end{align}
with the Green's function given by
\be\label{green}
g_p \left( \tau ,\, \tau_1 \right)  = \frac{1}{\Wr}2 i \,\Ima\left(\R_p(\tau_1)\R^*_p(\tau)\right).
\ee
To appreciate the generality of the problem, it is important to recognize that both expressions on the right-hand sides of Eq. \eqref{commutatorsgreen} do not depend \textit{in general} on the momentum $p$ in the super-horizon limit $p\tau,\,p\tau_1 \ll 1$, i.e.
\be
\left[ \Rh_{\bk}(  \tau_1) ,\,  \Rh_{\bp}(  \tau) \right]' =
\Wr \, g_p \left( \tau ,\, \tau_1 \right) \propto \Wr\tau_1,\quad 
 \left[ \Rh'_{\bk}(  \tau_1) ,\,  \Rh_{\bp}(  \tau) \right]' =
\partial_{\tau_1}
\left( \Wr g_p \left( \tau ,\, \tau_1 \right) \right) \simeq 
-\Wr.  
\label{commutatorsapprox}
\ee
Both relations originate from $ g_p(\tau, \tau_1) \propto \tau_1 $, which essentially arises from dimensional analysis. For illustrative purpose, let us consider two cases. 
The first is a quasi De Sitter evolution between $\tau_1$ and $\tau$, i.e. standard slow-roll (SR). The second is given by considering a phase of ultra-slow-roll (USR), i.e. $\eta = -6$, starting at $\tau_s$ and followed at $\tau_e$ by a phase of slow-roll. Then we take $\tau_1 \in[\tau_s,\tau_e]$ and $\tau > \tau_e$. Explicit computations of the Green's functions at leading order in \( p\tau_1 \) and \( p\tau \) give respectively:
\be
g^{\rm SR}_p(\tau,\tau_1) = \frac{\tau_1}{3 }  
\left(\left(\frac{\tau}{\tau_1}\right)^3 -1\right),\qquad
g^{\rm USR}_p(\tau,\tau_1)=\frac{\tau_1}{3}\left(1 - 2\left(\frac{\tau_1}{\tau_e}\right)^3 +\frac{(\tau\tau_1)^3}{\tau_e^6}  \right).
\ee
We will now show how the commutator relations in Eqs. \eqref{commutatorsapprox} enter into one-loop diagrams and lead to relative scale-invariant corrections to the power spectrum independently from considering or not a non-slow-roll phase.

\subsection{
Relative scale-invariant one-loop corrections}

Let us consider the one-loop contribution to the two-point function arising from two cubic interactions. From Eq. \eqref{fullinin}, we have:
\begin{equation}\label{twonested}
\langle\hat{\R}_{\bp}(\tau)\hat{\R}_{\bp'}(\tau)\rangle_{\Ha^{(3)}}=-\int_{\tauf}^{\tau}d\tau_{1}\int_{\tauf}^{\tau_{1}}d\tau_{2}\langle [\Ha^{(3)}(\tau_{2}),[\Ha^{(3)}(\tau_{1}),\hat{\R}_{\bp}(\tau)\hat{\R}_{\bp'}(\tau)]]\rangle.
\end{equation}
In general, an operator in the cubic Hamiltonian can be written as
\be\label{operatorcubic}
\Ha^{(3)} \supset   \lam(\tau)\cdot\R^A\R^B\R^C   , \quad \R^X \equiv \{\R,\R',\partial_i\R\},
\ee
where $\lam$ denotes a time dependent coupling and sometimes we found convenient to single out a factor proportional to the inverse of the Wronskian given in Eq. \eqref{Wronskian}, i.e.
\be\label{lam}
\lam \equiv \tilde{\lam} i \mathcal{W}^{-1}_{\tau} =\tilde{\lam}\, 2 a^2\epsilon . 
\ee
Unfolding the two-nested commutators in Eq. \eqref{twonested}, always results in two pairs of fields commutators along with four fields that will be Wick contracted in pairs.
One thus encounters two options, leading to two different sets of diagrams. The first possibility is when the two external $\Rh_{\bp}$ are both eaten by the two commutators:

\begin{align}\label{splitdiagram1}
\langle\hat{\R}_{\bp}\hat{\R}_{\bp'}\rangle_{1}=&-\int d\tau_1\int^{\tau_1} d\tau_2\int d\bK \lam(\tau_1)    \left[\Rh^A_1,\Rh_{\bp}\right]  \cdot   \left[ \Rh^A_{2},  \Rh_{\bp}\right]  \langle\Rh^B_{2}  \Rh^C_{2}\Rh^B_1\Rh^C_1\rangle,
\end{align}
and the second alternative is when only one of the two commmutators contains an external leg:\footnote{That is what has been labeled, in \cite{Senatore:2009cf}, as cut-in-the-middle and cut-in-the-side diagrams, respectively. This is how Feynman diagrams appear if you join legs only when two operators are contracted via a commutator.}
\begin{align}\label{splitdiagram2}
\langle\hat{\R}_{\bp}\hat{\R}_{\bp'}\rangle_{2}=
-\int d\tau_1\int^{\tau_1} d\tau_2\int d\bK \lam(\tau_1)    \left[\Rh^A_1,\Rh_{\bp}\right]  \cdot   \left[ \Rh^A_{2},  \Rh^B_1 \right]  \langle\Rh^B_{2}  \Rh^C_{2}\Rh^C_1 \Rh_{\bp}\rangle,
\end{align}
where to streamline notation and emphasize the commutation structure, we assign the same index to operators originating from the same vertex, as in Eqs. \eqref{compact}-\eqref{bigK}.

From the commutators in the limit $p\tau,p\tau_1 \ll 1$ in Eqs. \eqref{commutatorsapprox}, one notes that Eq. \eqref{splitdiagram1} becomes independent of the external momentum $p$ (at leading order), regardless of whether $\R^A_1 = \R_1$ or $\R^A_1 = \R'_1$. On the contrary, the option in Eq. \eqref{splitdiagram2} (where only one external leg have been eaten by the commutators) retains a $\bp$-dependence from the external leg outside the commutator. For instance, if at least one field operator has no derivative, i.e. $\R_2^B \equiv \R_2 $, from the Wick contractions one has 
$\wick[offset=1.03em,sep=0.4em]{\c1{\hat{\R}_2} \c1{\hat{\R}_{\bp'}}}  =  \delta(\bp + \bp') (2\pi)^3 |\R_p|^ 2$. 

    In summary, we have $\llangle\hat{\R}_{\bp}\hat{\R}_{\bp'}\rrangle_{1} \propto p^0$ and $\llangle\hat{\R}_{\bp}\hat{\R}_{\bp'}\rrangle_{2} \propto |\R_p|^2 $, the latter $\propto p^{-3}$ for a scale-invariant tree level power spectrum and where we use the double bracket notation to strip away the Dirac delta and the $(2\pi)^3$ factor, as specified in Eq. \eqref{defdoublebracket}.

The two contributions to the dimensionless power spectrum are thus given by:
\be\label{volume}
\p_{\R}^{\mathrm{1-loop},\,1}(p)=\frac{p^3}{2\pi^2}\llangle \Rh_{\bp}\Rh_{\bp'}\rrangle_1 \propto  O\left(\frac{p^3}{k^3}\right),
\ee
and 
\be\label{scaleinvariant}
\p_{\R}^{\mathrm{1-loop},\,2}(p)=\frac{p^3}{2\pi^2}\llangle \Rh_{\bp}\Rh_{\bp'}\rrangle_2 = \p^{\mathrm{tree}}(p)\int d\ln k \, C(k)\,
\ee
where we combined $p^3/2\pi^2$ and $|\R_p|^2$ from the Wick contraction into the dimensionless tree-level power spectrum (given in Eq. \eqref{treelevel}) and $C(k)$ is a dimensionless quantity that depends only on the short modes.
Contributions like the one in Eq. \eqref{scaleinvariant} are what we denote as relative scale-invariant corrections. Those arise, in general, if at least one operator in the cubic interaction \eqref{operatorcubic} has no derivative and because the Green's function, i.e. the first commutator in Eq. \eqref{splitdiagram2}, does not depend on the external momentum $p$ in the limit that this latter is well outside the horizon when the interaction is considered. For instance, corrections of the type \eqref{scaleinvariant} would appear whenever operator of the form $\Hamt \propto \R'\R^2,\,\R'^2\R,\,(\partial\R)^2\R, \mathrm{etc.}$ are used to make a one-loop diagram. Analogous reasoning shows the appearance of these types of corrections from loop diagrams involving a single quartic interaction. 

From now on we only consider contributions as the one in Eq. \eqref{splitdiagram2}, and in particular as in Eq. \eqref{scaleinvariant}. The contributions in Eq. \eqref{splitdiagram1} (and Eq.\eqref{volume}) are volume suppressed in the limit we are focusing here, i.e. the limit in which the momenta running in the loops are much shorter than the external ones.

\section{One-loop corrections as integral of three-point functions}\label{sec:oneloopthree}

Let us return to the one-loop diagrams constructed from two third-order Hamiltonians, as in Eq. \eqref{twonested}. We express a cubic interaction from the first nested Hamiltonian $\Ha^{(3)}(\tau_1)$ in the generic form given by Eq. \eqref{operatorcubic}. The second nested Hamiltonian is left implicit. The one-loop contribution can then be reorganized as follows:
\begin{align}\nonumber
\langle\hat{\R}_{\bp}\hat{\R}_{\bp'}\rangle=&-\int d\tau_1\int^{\tau_1} d\tau_2\int d\bK_1    \langle  [\Ha^{(3)}(\tau_2),[\lam\Rh^A_1\Rh^B_1\Rh^C_1,  \Rh_{\bp}\Rh_{\bp'}]] \rangle \\[1mm]
\label{loopthreepoint}
=&\int d\tau_1\int d\bK_1   \cdot i  [\Rh^A_1,\Rh_{\bp'}] \cdot \lam(\tau_1) \,\left( i\int^{\tau_1} d\tau_2 \langle  [ \Ha^{(3)}(\tau_2),  \Rh_1^B\Rh_1^C\Rh_{\bp}]  \rangle\right),
\end{align}
where $d\bK_1$ follows the notation in Eq. \eqref{bigK2} and permutations are understood. The term in parenthesis can be also expressed by using the notation
\be\label{firstderL}
\frac{\delta\Lt}{\delta \R^A}\Big|_{1} = -\lam \R_1^B\R_1^C.
\ee
For illustrative purpose, let us start by taking an operator without any derivative:\footnote{In truth, as we will see explicitly and as it has been pointed out already in the literature  \cite{Pimentel:2012tw}, the cubic action can be always written in such a way that there are no operators with no derivative or with just a single time derivative.} 
\be\label{toyexample}
\Ha^{(3)}(\tau_1) \propto \R_1^A\R_1^B\R_1^C = \R^3.
\ee
Now, since $\R^B\R^C \equiv \R^2$ in Eq. \eqref{loopthreepoint}, one immediately recognizes, in the parenthesis highlighted in Eq. \eqref{loopthreepoint}, the standard in-in formula for the three-point function $\langle \R_1 \R_1 \R_{\bp} \rangle$ in the commutator form. Although one consideration is in order. 
The``external legs" of this three-point function are two field operators borrowed from $\Ha^{(3)}(\tau_1)$ plus one external leg $\Rh_{\bp}$ from the full one-loop diagram. For the leading order (relative scale-invariant) loop corrections we are interested in, this latter leg always has to be taken outside the commutator making the three-point function---see previous section---and Wick contracted with one operator inside $\Ha^{(3)}(\tau_2)$. At that stage, since we are considering $p$ well outside the horizon at the times $\tau_i$ of the interactions, the time dependence of $\Rh_{\bp}$ becomes irrelevant, i.e., $\langle \Rh_2 \Rh_{\bp} \rangle \propto \mathrm{Re}\left(\zeta_p(\tau_2) \zeta^*_p(\tau)\right) \simeq |\R_p|^2$. We are therefore justified in approximating $\Rh_{\bp}(\tau) \simeq \Rh_{\bp}(\tau_1)$ in the term within parentheses in Eq. \eqref{loopthreepoint}, and in treating the three-point function as being evaluated at the time $\tau_1$, corresponding to the first nested Hamiltonian.

Let us make one further point that will aid in subsequent discussions.
The three-point function appearing in \eqref{loopthreepoint} should be considered as the genuinely connected one. The disconnected part would involve contracting two operators from the same interacting Hamiltonian $\Ha^{(3)} (\tau_2)$. That results in an overall non-1PI (non-one-particle-irreducible) diagram whose cancellation is usually ensured by adding a linear tadpole counterterm \cite{Cheung:2007st,Pimentel:2012tw}-- see discussion in section \ref{sec:tad}. 
This clarification turns out to be useful when adding diagrams from quartic induced Hamiltonian. 

For completeness of this illustrative case, let us write the one-loop correction to the dimensionless power spectrum arising from Eq. \eqref{loopthreepoint} augmented with an operator of the form \eqref{toyexample}:
\begin{align}
\p_{\R}^{\mathrm{1-loop}}(p) =& -\int^{\tau} d\tau_1\,\tilde{\lam} \,\Wr^{-1}  [\Rh_{\bk}(\tau_1),\Rh_{\bp}(\tau)]' \int \frac{d\bk}{(2\pi)^3} \frac{p^3}{2\pi^2}\llangle \Rh_{\bk}\Rh_{\bk}\Rh_{\bp} \rrangle \\[1mm]
 =&-\int^{\tau} d\tau_1\,\tilde{\lam} \, g_{p}(\tau,\tau_1) \int d\ln k\,\frac{k^3}{2\pi^2} \frac{p^3}{2\pi^2} \llangle \Rh_{\bk}\Rh_{\bk}\Rh_{\bp} \rrangle,
\end{align}
where we write the coupling as in Eq. \eqref{lam} and the commutator as expressed in Eq. \eqref{commutatorsgreen}.

\subsection{Correlators with time derivatives and quartic induced Hamiltonian}\label{timederandquartic}

\subsubsection*{Correlators with time derivatives}
The situation becomes more involved in the presence of time-derivative operators in \(\Ha^{(3)}(\tau_1)\). In particular, if \(\Rh^B_1 \Rh^C_1\) in Eqs.\ \eqref{loopthreepoint} (and \eqref{firstderL}) contains time derivatives, then the term in parentheses would not constitute a three-point function on its own.
To illustrate this clearly, let us write explicitly correlators involving fields with one and two time derivatives\footnote{As is customary throughout this manuscript, with a slight abuse of notation, fields on the right-hand side are understood to be in the interaction picture.}:
\begin{align}\nonumber
\langle \Rh'_{\bk} \Rh_{\bk'}  \Rh_{\bp} \rangle +\langle \Rh_{\bk} \Rh'_{\bk'}  \Rh_{\bp} \rangle  =&
i\int_{\tauf}^{\tau_1}d\tau_{2}\langle [\Ha^{(3)}(\tau_{2}),\left(\hat{\R}'_{\bk}(\tau_1) \hat{\R}_{\bk'}(\tau_1)+\hat{\R}_{\bk}(\tau_1) \hat{\R}'_{\bk'}(\tau_1)\right)\hat{\R}_{\bp}]\rangle\\[1mm] \label{correlatoronederivative}
+ i \langle [\Ha^{(3)}(\tau_1)&,\hat{\R}_{\bk}(\tau_1)] \hat{\R}_{\bk'}(\tau_1)\hat{\R}_{\bp}(\tau_1) \rangle + i \langle \hat{\R}_{\bk}  [\Ha^{(3)}(\tau_1),\hat{\R}_{\bk'}(\tau_1)]\hat{\R}_{\bp}(\tau) \rangle,
\end{align}
and
\begin{align}\nonumber
\langle \Rh'_{\bk} \Rh'_{\bk'}  \Rh_{\bp} \rangle =&
i\int_{\tauf}^{\tau_1}d\tau_{2}\langle [\Ha^{(3)}(\tau_{2}),\hat{\R}'_{\bk}(\tau_1) \hat{\R}'_{\bk'}(\tau_1)\hat{\R}_{\bp}(\tau_1)]\rangle \\[1mm] \label{correlatortwoderivative} &+ i \langle [\Ha^{(3)}(\tau_1),\hat{\R}_{\bk}(\tau_1)] \hat{\R}'_{\bk}(\tau_1)\hat{\R}_{\bp} \rangle +  i \langle \hat{\R}'_{\bk}(\tau_1) [\Ha^{(3)}(\tau_1),\hat{\R}_{\bk}(\tau_1)] \hat{\R}_{\bp} \rangle.
\end{align}
The expressions above can be derived by writing explicitly the time evolution of operators in the interaction picture. Restoring, just for this purpose, the (omitted) $I$ on interacting fields, one may insert on the left-hand side of Eqs. \eqref{correlatoronederivative}-\eqref{correlatortwoderivative}:
\be
\Rh(\tau_1) = \mathcal{U}_{\tau_1}^{\dagger}\Rh_{\rm I}\mathcal{U}_{\tau_1},\qquad \mathcal{U}_{\tau_1} \equiv T\left(e^{-i\int_{\tauf}^{\tau_1}d\tau'\Ha_{I}(\tau')}\right)
\ee
where $\mathcal{U}_{\tau_1} $ is the unitary operator evolving the field in the interaction picture. By using $\partial_{\tau_1}\mathcal{U}_{\tau_1} = -i\mathcal{U}_{\tau_1}H_{\rm I}$ and expanding in perturbation theory, one obtains the standard in-in formula (first lines) plus the contact terms (second lines) in Eqs. \eqref{correlatoronederivative}-\eqref{correlatortwoderivative}. 
Since the contact terms involve equal-time commutators, they are nonzero only if the Hamiltonian considered contains operator with time derivatives.

As briefly reviewed here and as first pointed out by the authors in \cite{Pimentel:2012tw}, it is only by summing diagrams from the quartic induced Hamiltonian—which we will define shortly—that we can account for the contact terms on the second lines of Eqs. \eqref{correlatoronederivative} and \eqref{correlatortwoderivative}.
 Therefore, it is only by adding these diagrams that we can express the full one-loop corrections as an integral involving three-point functions.

\subsubsection*{Quartic induced Hamiltonian}

When defining the interacting Hamiltonian in perturbation theory, the conjugate momentum becomes non-linearly related to the field $\R$ and its first derivative $\R'$. That ultimately results in a quartic contribution to the Hamiltonian.  
We refer to this as the \textit{quartic induced Hamiltonian} and label it as $\mathcal{H}^{(4)}_{3}$. In Appendix \ref{appquarticinduced}, we show how it can be derived from the cubic Lagrangian $\Lt$ using the following relation:
\be\label{masterquarticinduced}
\mathcal{H}^{(4)}_{3} = \frac{1}{2 (2 a^2 \epsilon) }\left(\frac{\delta \mathcal{L}^{(3)}}{\delta \zeta'}\right)^2.
\ee
One-loop corrections to the power spectrum from these quartic contributions can be written as\footnote{To simplify notation we omit the explicit $\int d^3 x$ in passing from the Hamiltonian to the Hamiltonian density, it will be automatically accounted for when moving to Fourier space.
}
\begin{align}\nonumber
\langle\Rh_{\bp}\Rh_{\bp'}\rangle_{H^{(4)}_3} &\equiv i \int^{\tau} d\tau_1 \langle[H^{(4)}_3, \Rh_{\bp}\Rh_{\bp'}]\rangle= i \int^{\tau} d\tau_1 \laa\left[ \frac{1}{2(2a^2\epsilon)}\frac{\delta \Lt}{\delta \R'}\frac{\delta \Lt}{\delta \R'}, \Rh_{\bp}\Rh_{\bp'}\right]\raa \\[1.1mm] \label{quarticinducedoneloop}
&=
 \int^{\tau}d\tau_1\left(\frac{i}{2 a^2 \epsilon}\laa \frac{\delta \Lt}{\delta\R'} \left[ \frac{\delta \Lt}{\delta\R'},\Rh_{\bp}\right]\Rh_{\bp'}\raa  + \frac{i}{2 a^2 \epsilon}\laa \left[ \frac{\delta \Lt}{\delta\R'},\Rh_{\bp}\right] \frac{\delta \Lt}{\delta\R'} \Rh_{\bp'}\raa\right).
 \end{align}
Each of the two terms in the last expression can be further rearranged by taking two steps. First, we rewrite the commutators in Eq. \eqref{quarticinducedoneloop} as 
 \begin{align}\label{commextra}
 \left[ \frac{\delta \Lt}{\delta\R'},\Rh_{\bp}\right] = \frac{\delta \Lt}{\delta\R^A\delta\R'}\cdot [\Rh^A,\R_{\bp'}].
\end{align}
Second, the operator (in Fourier space) \( \frac{i}{2a^2\epsilon}\frac{\delta\Lt}{\delta\R'} \) outside each commutator in the second line of Eq. \eqref{quarticinducedoneloop} can be replaced using the following equal-time commutator:
\be
[\Ha^{(3)}(\tau_1),\Rh_{\bk}(\tau_1)] = -\frac{\delta\Lt}{\delta\R'}\Big|_{\bk}[\Rh'_{\bk},\Rh_{\bk}]' =-\frac{\delta\Lt}{\delta\R'}\Big|_{\bk}(-\Wr) = \frac{i}{2 a^2\epsilon } \frac{\delta\Lt}{\delta\R'}\Big|_{\bk},
\ee
where we recognize the Wronskian specified in Eq. \eqref{commutator01}.

Going to Fourier space, one can express the one-loop contribution from the quartic induced Hamiltonian in Eq. \eqref{quarticinducedoneloop} as\footnote{Note that if $\delta\Lt/\delta\R'\equiv \lam \R^A\R^B$, $\delta\Lt/\delta\R'|_{\bk} =\lam \int \frac{d\bk'}{(2\pi)^3} \Rh^A_{\bk'}\Rh^B_{\bk-\bk'}$.}
\begin{align}\nonumber
\langle\Rh_{\bp}\Rh_{\bp'}\rangle_{H^{(4)}_3}=&   -\int^{\tau} d\tau_1 \int d\bK_1 \,i  [\Rh_1^A,\Rh_{\bp'}]\cdot\left(
i\laa [\Ha^{(3)}(\tau_1),\Rh_1(\tau_1)]  \,\frac{\delta \Lt}{\delta\R^A\delta\R'}\Big|_1\, \Rh_{\bp}\raa \right.  \\[1mm] \label{quarticinducedfinal}
&\left.  + i\laa   \,\frac{\delta \Lt}{\delta\R^A\delta\R'}\Big|_1 [\Ha^{(3)}(\tau_1),\Rh_1(\tau_1)]\, \Rh_{\bp}\raa       \right),
\end{align}
where the subscript one, here redundant, labels as before operators coming from the same vertex (and so evaluated at the same time).
\subsection{General formula}
Within the parentheses of the expression in Eq. \eqref{quarticinducedfinal} above, one can recognize the type of contact terms found in Eqs. \eqref{correlatoronederivative}-\eqref{correlatortwoderivative}, where the role of the two short modes operators $\R^A_{\bk}\R^B_{\bk'}$ is now represented by $\delta \Lt / \delta \R^A$.
Therefore, by rewriting the cubic one-loop diagrams in Eq. \eqref{loopthreepoint}, using Eq. \eqref{firstderL}, and summing them with the quartic induced diagrams from Eq. \eqref{quarticinducedfinal} above, one arrives at the simple and general formula \cite{Pimentel:2012tw}:
\begin{align}\label{finalformula}
\langle\hat{\R}_{\bp}\hat{\R}_{\bp'}\rangle=\langle\Rh_{\bp}\Rh_{\bp'}\rangle_{H^{(3)}}+
\langle\Rh_{\bp}\Rh_{\bp'}\rangle_{H^{(4)}_3}
=
-\int d\tau_1\int d\bK_1  \cdot i   [\Rh^A_1,\Rh_{\bp'}] \cdot  \,\laa \frac{\delta\Lt}{\delta \R^A}\Big|_{1}\Rh_{\bp}  \raa.
\end{align}
Using the commutator-Green's functions expressions in Eqs. \eqref{commutatorsgreen}-\eqref{commutatorsapprox}, 
one can write the analogous contribution to the dimensionless power spectrum as:
\begin{align}\nonumber
\p_{\R}^{\mathrm{1-loop}}(p) =&\int^{\tau} d\tau_1\,i \Wr \,  \int d\ln k\,\frac{k^3}{2\pi^2} \frac{p^3}{2\pi^2} \llangleb \frac{\delta\Lt}{\delta \R}\Big|_{\bk,\bk'}\Rh_{\bp} \rrangleb\\[1mm]
 &-\int^{\tau} d\tau_1\,i \Wr \, g_{p}(\tau,\tau_1) \int d\ln k\,\frac{k^3}{2\pi^2} \frac{p^3}{2\pi^2} \llangleb \frac{\delta\Lt}{\delta \R'}\Big|_{\bk,\bk'}\Rh_{\bp} \rrangleb.
\end{align}
We have thus shown how to express, in general, the one-loop corrections to the power spectrum as integrals of three-point functions. We now remind the reader that our ultimate goal is to apply the consistency relations on the three-point functions (that we derive in Sec. \ref{sec:consistency}) in order to rewrite all relative scale-invariant one-loop corrections for a transient non-slow-roll phase (which we compute in Sec. \ref{sec:explicitoneloop}) as a total derivative with respect to the looping momenta.

\subsection{Quadratic induced Hamiltonian from tadpoles counterterms}\label{sec:tad}
Let us conclude this section with a technical remark.
In order for the consistency relations to be sufficient to conclude regarding the fate of the one-loop corrections from short to the long-wavelength modes, we found it necessary to augment the argument in \cite{Pimentel:2012tw} by explicitly mentioning one additional ingredient. 
In particular, direct computations reveal that the quartic induced Hamiltonian introduces additional contributions beyond those required to use the consistency relations. 

To trace the origin of these contributions, consider, for example, a three-point correlator with one $\Rh'$ inside the one-loop formula in Eq. \eqref{finalformula}, as given by Eq. \eqref{correlatoronederivative}.
First, let us focus on the first line of Eq. \eqref{correlatoronederivative}. The connected part of this latter, i.e. no contraction among operators in $\Ha^{(3)}(\tau_2)$, is the part entering in the connected 1-PI one-loop diagram, as mentioned below Eq. \eqref{loopthreepoint}. The disconnected part would instead lead to non-1-PI one-loop diagrams. Contributions of this type are taken care of by adding linear tadpole counterterms. Schematically we can write them in the following form:
\be\label{tadL}
\Lag_{\rm tad}^{(1)} = c \,\R' + \tilde{c}\R,\qquad c =- \laa \frac{\delta\Lt}{\delta\R'}\raa,\quad \tilde{c}=- \laa \frac{\delta\Lt}{\delta\R}\raa,
\ee
where the braket defying $c,\tilde{c}$ denotes the Wick contraction of the two field operators in $\frac{\delta\Lt}{\delta\R'},  \frac{\delta\Lt}{\delta\R}$ respectively, and again we omit spatial derivatives, as they play no role here.
This would ensure the cancellation of all 1-non-PI diagrams in which two legs from the same cubic interaction are contracted.

Let us now turn to the second line in Eq. \eqref{correlatoronederivative}. We highlight two types of contributions from these terms. The first type involves cases where we \textit{do not} allow for contractions between the two operators outside the equal-time commutators. These are the terms that once combined with the connected part of the first line in Eq. \eqref{correlatoronederivative} lead to the connected three-point functions inside the loop integral. The second type includes cases where contractions between the two operators outside the equal-time commutators are permitted. These latter terms produce the excess contributions that are the focus of our concern here.

In general, these contributions arise from one-loop diagrams built with quartic-induced Hamiltonian interactions, where two operators from the same \(\delta \mathcal{L}_t / \delta \mathcal{R}'\) in the decomposition on the first line of Eq. \eqref{quarticinducedoneloop} are contracted, i.e.,

\begin{align}\label{termcanc}
\langle\Rh_{\bp}\Rh_{\bp'}\rangle_{H^{(4)}_3} &\supset 2\cdot i  \int^{\tau} d\tau_1 \laa\left[ \frac{1}{2(2a^2\epsilon)}\laa\frac{\delta \Lt}{\delta \R'}\raa\frac{\delta \Lt}{\delta \R'}, \Rh_{\bp}\Rh_{\bp'}\right]\raa.
\end{align}
Interestingly, those additional diagrams cancel exactly with peculiar contributions coming from 
 the “quadratic induced Hamiltonian” from the tadpole counterterms. In the same way that a cubic Lagrangian induces a quartic Hamiltonian, we found that the linear tadpole Lagrangian in Eq. \eqref{tadL} induces a quadratic Hamiltonian--see Appendix \ref{appquarticinduced}. This is given by
 
\be\label{quadratictadpole}
\Ham^{(2)}_{1} = \frac{c}{2 a^2\epsilon  }\, \frac{\delta\Lt}{\delta\R'},
\ee
where $c$ is defined in Eq. \eqref{tadL}. This provides the following contributions to the two-point correlator:
\be
\langle\Rh_{\bp}\Rh_{\bp'}\rangle_{H^{(2)}_1} \equiv i \int^{\tau} d\tau_1 \langle[H^{(2)}_1, \Rh_{\bp}\Rh_{\bp'}]\rangle = -  \frac{i}{2 a^2\epsilon  } \int^{\tau} d\tau_1 \laa\left[\laa \frac{\delta\Lt}{\delta\R'}\raa  \frac{\delta\Lt}{\delta\R'} , \Rh_{\bp}\Rh_{\bp'}\right]\raa.
\ee
These diagrams precisely cancel the ones in Eq. \eqref{termcanc}. Therefore, when performing explicit one-loop computations with the quartic induced Hamiltonian, we will also include contributions from the Hamiltonian \eqref{quadratictadpole}.\footnote{In practice, under the long-wavelength limit (\(p \ll k\)), these spurious contributions, whose cancellation is ensured by \eqref{quadratictadpole}, would be included only if, in the commutator in the second line of Eq. \eqref{quarticinducedoneloop}, the operator \(\frac{\delta \Lt}{\delta\R^A \delta\R'}\),  as specified in Eq. \eqref{commextra}, is proportional to \(\R\) (specifically, the diagrams derived from Eq. \eqref{correlatoronederivative}). For instance, if \(\frac{\delta \Lt}{\delta\R^A \delta\R'} \propto \R'\), then contributions like those in Eq. \eqref{termcanc} are proportional to \(|\R'_{\bp}|^2\) and are thus already suppressed.
}

\section{(Relevant part of) The Hamiltonian}\label{sec:Hamiltonian}

In this section, we single out the part of the Hamiltonian relevant for computing the one-loop power spectrum in the long-wavelength limit, at first order in $\epsilon$ and for arbitrary $\eta$.

We start by considering standard single-field inflation within the 
``comoving gauge'', where field fluctuations are set to zero. In this gauge, the spatial component of the metric is expressed as $g_{ij}= a^2 e^{2\R}$, where $\R$ is the so-called comoving curvature perturbation. This is the sole degree of freedom required to characterize the dynamics of the scalar fluctuations. Upon expanding the single-field action and integrating the constraints, the Lagrangian density, defined from the action as $S=\int d\tau d \bx \,\mathcal{L}$,  can be written at second-order in $\R$ as follows ($\Mp=1$)
\begin{align}
\label{quadraticR} 
\mathcal{L}^{(2)}&=\frac{1}{2}  (2 a^2 \epsilon) \left[ \R'^2 - (\partial_i\R)^2 \right]. 
\end{align}
The third-order Lagrangian can be expressed as detailed below \cite{Maldacena:2002vr}:
\be\label{fullthird}
\Lt =  \mathcal{L}^{(3)}_{\mathrm{bulk}} + \mathcal{L}^{(3)}_{\partial} + \mathcal{L}^{(3)}_{\mathrm{eom}}
\ee
where $\mathcal{L}^{(3)}_{\partial}$ label total time derivative terms, generally referred to as boundary terms, $\mathcal{L}^{(3)}_{\mathrm{eom}}$ denotes terms proportional to the linear equations of motion (eom), and $\mathcal{L}^{(3)}_{\mathrm{bulk}}$ is what is left as the standard ``bulk Lagrangian".
In particular, we decide to arrange the three contributions as follows \cite{Maldacena:2002vr,Arroja:2011yj,Garcia-Saenz:2019njm}: 
\begin{align}\nonumber
\mathcal{L}^{(3)}_{\mathrm{bulk}} =\,\, &   a^2 \left(      \epsilon( \epsilon - \eta) \R\R'^2 +  \epsilon(\epsilon + \eta)   \R(\partial \R)^2   
-2 \epsilon^2 \R' \partial_i\R\partial_i\partial^{-2}\R'\right)\\[1mm] \label{third_zeta_bulk}
&+   a^2\frac{\epsilon^3}{4}(\partial_i\partial^{-2}\R')(2\,\R'\partial_i\R + \partial^2\R\partial_i\partial^{-2}\R'),
\end{align}
then we define
\be
\mathcal{L}^{(3)}_{\partial} \equiv   \frac{d}{d \tau} \widetilde{\mathcal{L}}^{(3)}_{\partial},
\ee
with
\begin{align}\nonumber
\widetilde{\mathcal{L}}^{(3)}_{\partial}=&  -\frac{\epsilon a^2}{ a H} \R'^2  \R + \frac{\epsilon  a^2}{2 
 a^2 H^2} \R'( \partial_i\partial_j \partial^{-2}(\R \partial_i\partial_j \R) -\R\partial^2\R )+ \frac{\epsilon^2 a^2}{2 a H}\left(\R\R'^2 - \R (\partial_i\partial_j\partial^{-2}\R')^2\right),\\[1mm] \label{boundary}
 & -9 a^3 H \R^3 - \frac{a^2}{4(a H)^3}(\partial\R)^2(\partial^2\R) +\frac{a^2}{a H}(1-\epsilon)\R(\partial\R)^2,
\end{align}
where we keep factors of $(a H)$ explicit, and finally 
\be\label{eom}
\mathcal{L}_{\mathrm{eom}}^{(3)} =    \mathcal{E}_\R   \left[\frac{2}{a H}\R'\R +\frac{1}{2 a^2 H^2}\left(\partial^2\R \R -\partial^{-2}\partial_i\partial_j(\R\partial_i\partial_j\R)\right) -  \frac{\epsilon}{a H} (\R\R' -\partial_i\partial_j\partial^{-2}(\R\partial_i\partial_j \partial^{-2}\R') \right] 
\ee
with 
\be\label{lineareom}
\mathcal{E}_\R \equiv -\frac{1}{2 }\frac{\delta \mathcal{L}^{(2)}}{\delta \R} = (a^ 2 \epsilon \mathcal{\R}')' - a^ 2 \epsilon  \partial^2\R.
\ee
The historical motivations for these types of rewritings stem from computations of tree-level three-point functions (bispectra). As written in the explicit form above\footnote{One could also, as we do in \eqref{firstlag}, reorganize the first two terms in Eq. \eqref{third_zeta_bulk} so that a term proportional to $\eta'$ and $\eta$ appears in $\mathcal{L}^{(3)}_{\mathrm{bulk}}$ and $\widetilde{\mathcal{L}}^{(3)}_{\partial}$ respectively.}, the bulk action effectively reveals the true magnitude of contact diagrams on super-horizon scales, telling us that they start at second order in the slow-roll parameters.
This is not necessary the case for one-loop diagrams, as we will discuss. 

Let us now have a closer look to the various terms in Eq. \eqref{fullthird} and clarify the ones needed for our purposes. Although our main argument to prove the absence of one-loop corrections from short modes to long modes does not rely on any slow-roll ($\epsilon$) truncation, we show that explicitly by truncating the action at first order in $\epsilon$. This is also where the relevant contributions for the non-slow-roll phases discussed here lie.\footnote{To extend the explicit proof in the next sections by including terms at order $\epsilon^2$ from the action, one should also compute the tree-level power spectrum for a non-slow-roll transition at next to leading order in $\epsilon$. We may leave this for the future.} 

Thus, we consider the following terms in the Lagrangian:
\begin{itemize}
\item
From the \textit{bulk Lagrangian} in Eq. \eqref{third_zeta_bulk} we keep the two operators proportional to $\eta$ from the first two terms. That is the leading order in $\epsilon$.
\item
From the \textit{boundary terms Lagrangian}, we need to retain only the first term in Eq. \eqref{boundary}. The second and third terms provide a suppressed contribution to the squeezed bispectra (as already pointed out in \cite{Motohashi:2023syh}). The contributions of these latter terms to the one-loop power spectrum of infrared modes can therefore be disregarded following the findings of the previous section.\footnote{Strictly speaking, that is not immediately guaranteed when an operator giving negligible contribution to the squeezed bispectra enters in the formula in Eq. \eqref{twonested} as $\Hamt(\tau_1)$. However, even if not zero, these diagrams will sum up to three-point functions computed from the cubic Hamiltonian singled out in this section. It is therefore consistent to neglect them for our purposes.} The second line in Eq. \eqref{boundary} consists of ``boundary terms without time derivatives". It is (well) known that these have no effect on the correlators at any order in perturbation theory (see, for instance, \cite{Burrage:2011hd}, where the path integral formalism has been used for this purpose and \cite{Braglia:2024zsl}). Using the operator formalism (employed throughout this paper), it is straightforward to show this for contact diagrams, using the nested commutator formula \eqref{fullinin} and the fact that fields commute at equal time. However, the prove is less trivial when considering loops. In Appendix \ref{APPboundarynoder}, we demonstrate in general, that one-loop corrections from cubic boundary interactions without time derivatives cancel exactly with the contributions from their quartic induced interactions--see also \cite{Braglia:2024zsl} for an alternative proof.
\item
Finally, the \textit{equation of motion terms} do not
contribute at any order to correlators computed within the interaction picture, where fields are evaluated on the linear equation of motion. However, these terms are crucial for accurately computing the quartic Hamiltonian induced by the cubic Lagrangian (see section \ref{timederandquartic}). To achieve this, one must differentiate the cubic Lagrangian with respect to the first derivative of \(\zeta\). Including the equation of motion terms ensures that the Lagrangian, consistently, does not contain second derivatives of the field, depending only on \(\zeta\) and \(\zeta'\). We thus keep, for the purpose of computing the quartic Hamiltonian, the first term in Eq. \eqref{eom}. 
Note that we do not retain the second and third term in Eq. \eqref{eom}. The former because it does not bring any operator with first derivative of $\R$ once it is combined with the second term in Eq. \eqref{boundary}. The latter since it is second order in $\epsilon$.
\end{itemize}
Summarizing all points, from now on we will consider the following cubic Lagrangian:
\be\label{secondlag}
\Lt = -  \eta \epsilon a^2 \R\R'^2 +   \eta \epsilon a^2   \R(\partial \R)^2 +  \frac{d}{d\tau}\left[-\frac{\epsilon a^2}{ a H} \R'^2  \R\right] +   \mathcal{E}_\R  \frac{2}{a H}\R'\R.
\ee
By rewriting the first two terms in the previous expression, we can express the Lagrangian in the following equivalent form:
\be\label{firstlag}
\Lt  =   \,\frac{a^2 \epsilon}{2}\eta' \zeta^2 \zeta' -    \frac{d}{d\tau} \left[\frac{a^2 \epsilon \eta}{2}\zeta^2\R' +\frac{\epsilon a^2}{ a H} \R'^2  \R\right] +   \mathcal{E}_\R  \left( \frac{\eta}{2}\R^2 +\frac{2}{a H}\R'\R\right).
\ee
In the context discussed, different options have been considered. For instance, Ref. \cite{Kristiano:2022maq} considers the first operator in Eq. \eqref{firstlag}, while \cite{Firouzjahi:2023aum} takes into account the first two terms of Eq. \eqref{firstlag}. As already mentioned in \cite{Fumagalli:2023hpa}, using the first two terms in Eq. \eqref{secondlag} or the first bulk and boundary term in  Eq. \eqref{firstlag} provides with equivalent results. 

Further, by expanding the last two terms in Eq. \eqref{secondlag}, using standard relations on background quantities listed in Appendix \ref{backquant}, one can rewrite the Lagrangian as

\be\label{thirdlag}
\Lt =   \eta \epsilon a^2   \R(\partial \R)^2 -\left( \frac{\epsilon a^2}{aH}\R'^3  -3\epsilon a^2\R'^2\R + \frac{2 a^2\epsilon}{a H}\R'\R\partial^2 \R \right).
\ee
where we used
\be
\frac{d}{d\tau}\left[-\frac{\epsilon a^2}{ a H} \R'^2  \R\right] = -\frac{2}{a H}\R\R'(\epsilon a^2 \R')' -\frac{\epsilon a^2}{a H} \R'^3 + (\eta+3-\epsilon)\epsilon a^2 \R'^2\R,
\ee
and neglected terms at second order in $\epsilon$.
We stress that Eqs.\eqref{secondlag}-\eqref{firstlag}-\eqref{thirdlag} are completely equivalent way of expressing the same Lagrangian.
It turns out that Eqs. \eqref{secondlag}-\eqref{firstlag} are more convenient for computing in-in formulas, such as one-loop diagrams with two cubic vertices. In these cases, one can neglect the term proportional to the equations of motion and use boundary terms to simplify the time integration. We will demonstrate this explicitly in Sections \ref{sec:consistency} and \ref{sec:explicitoneloop}. Conversely, as mentioned earlier, for computing the quartic induced Hamiltonian (see next sections) and parts of three-point function in presence of time derivative (see Sec. \ref{squeezedderivativeoperator}), the explicit form in Eq. \eqref{thirdlag} will prove to be more convenient.

\subsection{Cubic Hamiltonian}

To set notation, we recall that the Hamiltonian $H$ is defined in terms of the Hamiltonian density $\Ham$ as follows:
\be
H = \int d^3 x \,\mathcal{H},
\ee
while the Hamiltonian density at third order is simply given by (see Appendix \ref{appquarticinduced}):
\be
\Hamt = -\Lt,
\ee
and we express it explicitly using the two equivalent forms of the Lagrangian in Eqs. \eqref{secondlag} and \eqref{firstlag}.

From Eq. \eqref{firstlag} we define:
\begin{align}\label{firstset} 
    \mathcal{H}^{(3)}&=\,\, \mathcal{H}^{(3)}_a +\mathcal{H}^{(3)}_b + \mathcal{H}^{(3)}_e\,\\[1mm] \nonumber
    \mathcal{H}^{(3)}_a &=-  \frac{a^2 \epsilon}{2}\eta' \zeta^2 \zeta',\quad 
    \mathcal{H}^{(3)}_b =   \,\frac{d}{d\tau} \left[\frac{a^2 \epsilon \eta}{2}\zeta^2\R'\right],\quad
    \mathcal{H}^{(3)}_e =  \frac{d}{d\tau} \left[\frac{a \epsilon}{H}\zeta\R'^2\right].
\end{align}

From Eq. \eqref{secondlag}, we can also rewrite the cubic Hamiltonian in the equivalent manner: 
\begin{align} \label{secondset}
\mathcal{H}^{(3)}&=\mathcal{H}_c^{(3)}+\mathcal{H}_d^{(3)} +\mathcal{H}_e^{(3)} \\[1mm] \nonumber
\mathcal{H}_c^{(3)} &=     a^2\epsilon \eta  \R'^2\R,\quad \mathcal{H}_d^{(3)} =\frac{1}{2}     a^2 \epsilon\eta \R^2 \partial^2\R, \qquad
    \mathcal{H}^{(3)}_e =    \frac{d}{d\tau} \left[\frac{a \epsilon}{H}\zeta\R'^2\right].
\end{align}
We will use the expression in Eq. \eqref{firstset} to compute the consistency relations in Sec. \ref{sec:consistency} and the form in Eq. \eqref{secondset} to compute one-loop diagrams. There we comment on the reshuffling needed if one starts with the set in Eqs. \eqref{firstset}.

\subsection{Quartic induced Hamiltonian from cubic Lagrangian}

As already discussed in Sec. \ref{timederandquartic} and detailed in Appendix \ref{appquarticinduced}, when defining the Hamiltonian in perturbation theory, the cubic Lagrangian induces a quartic contributions that can be written as:

\be\label{masterquarticinduced2}
\mathcal{H}^{(4)}_{3} = \frac{1}{2 (2 a^2 \epsilon) }\left(\frac{\delta \mathcal{L}^{(3)}}{\delta \zeta'}\right)^2.
\ee
To compute $\mathcal{H}^{(4)}_{3}$ from the cubic Lagrangian $\Lt$, we use the explicit form of this latter as given in Eq. \eqref{thirdlag}. This is well-suited for this purpose since the second time derivatives of \(\R\) are explicitly absent. We obtain

\begin{align}\label{quarticinducedset}
\mathcal{H}^{(4)}_{3} &=\,\, \sum_{X}   \mathcal{H}^{(4)}_{X},\qquad  X=\{A,B,C,D,E,F\},\\[1mm] \label{quarticinducedset1}
\mathcal{H}^{(4)}_{A} &= 9 \epsilon a^2 \R'^2\R^2,\qquad \mathcal{H}^{(4)}_{B} =\frac{\epsilon a^2}{(a H)^2}\R^2(\partial^2\R)^2,\qquad \mathcal{H}^{(4)}_{C} = - \frac{9\epsilon a^2}{a H}\R'^3\R,\\[1mm] \label{quarticinducedset2}
\mathcal{H}^{(4)}_{D} &= -\frac{6 \epsilon a^2}{a H}\R^2\R'\partial^2\R,\qquad 
\mathcal{H}^{(4)}_{E} = \frac{3\epsilon a^2}{(a H)^2}\R\R'^2\partial^2\R,\qquad \mathcal{H}^{(4)}_{F} = \frac{9\epsilon a^2}{4 (a H)^2}\R'^4.
\end{align}
Note that $\mathcal{H}^{(4)}_{F}$ give suppressed contribution ($\propto |\R'_p|^2$) to loop-diagrams with soft modes in the external legs. We will thus disregard it in the following. 
It is worth mentioning that there is no explicit dependence on $\eta$ in the previous expressions. That is simply due to the fact the Lagrangian \eqref{thirdlag} has, in truth, no $\eta$ dependent terms containing time derivatives. Nevertheless, to keep separate the contributions induced by the term explicitly dependent on $\eta$ in \eqref{secondset} and the one from $\Ham_{\rm e}$, it will turn useful to introduce:
\begin{align}\label{quarticinducedset3}
\mathcal{H}^{(4)}_{A} &= - 3\eta \epsilon a^2 \R'^2\R^2 + 3(3+\eta) \epsilon a^2 \R'^2\R^2  \equiv \mathcal{H}^{(4)}_{A_1} + \mathcal{H}^{(4)}_{A_2} \\[1mm] \label{quarticinducedset4}
\mathcal{H}^{(4)}_{D} &= \eta\frac{\epsilon a^2}{a H}\R^2\R'\partial^2\R  -(6+\eta)\frac{ \epsilon a^2}{a H}\R^2\R'\partial^2\R \equiv \mathcal{H}^{(4)}_{D_1} + \mathcal{H}^{(4)}_{D_2}.
\end{align}

\subsection{Quartic Hamiltonian from residual diffeomorphism invariance}\label{sec:diffinduced}
In the $\R$ gauge, the action has a residual diff. symmetry\footnote{We consider the parameters of the diff. transformation as time independent. That ensures constraints can be integrated in the standard way, i.e. by imposing boundary conditions that vanish at spatial infinity \cite{Urakawa:2009my,Urakawa:2010it}. The action for $\R$, after integrating the lapse and shift function, is then invariant under the transformation in Eq. \eqref{diff}. That is what is needed to infer the quartic interactions.} \cite{Urakawa:2009my,Urakawa:2010it}-\cite{Pimentel:2012tw}:
\be\label{diff}
\R\rightarrow \R + b,\quad x^i\rightarrow x^i e^{-b}+ C^i
\ee
for which the combination 
\be\label{block}
e^{-\R}\partial_i\R
\ee
is invariant.
Thus, given a cubic Lagrangian, one can infer the necessary quartic terms for ensuring diff invariance by constructing nonlinear blocks based on Eq. \eqref{block}. These blocks should reproduce the cubic terms when expanded, allowing the quartic terms to be identified from the subsequent order of the expansion, i.e. 
\begin{align}\label{firstdiffex}
\Lt &\supset -c_1\R (\partial_i\R)^2\, \implies\, \Lag_{\rm diff}^{(4)} \supset c_1\R^2 (\partial_i\R)^2 \\[1mm]
\Lt &\supset -c_1\R' (\partial_i\R)^2 \,\implies\, \Lag_{\rm diff}^{(4)} \supset 2 c_1\R'\R (\partial_i\R)^2 \\[1mm]
\Lt &\supset -c_1\R'\R \partial^2\R\,\implies\, \Lag_{\rm diff}^{(4)} \supset -2 c_1\R'\R (\partial_i\R)^2 -c_1\R^2\partial\R\partial\R'.
\end{align}
From the Lagrangian in the form \eqref{thirdlag}, we thus derive the relevant part of the quartic induced Hamiltonian: 
\begin{align}\label{quarticdiffset}
\Ham^{(4)}_{\mathrm{diff}} &=\sum_X \Ham^{(4)}_{\mathrm {diff},\,X}, \qquad  X=\{A,B,C\},\\[1mm] \label{quarticdiffset1}
\Ham^{(4)}_{\mathrm {diff},\,A} &=a^2 \epsilon \eta \,\R^2(\partial\R)^2,\qquad \Ham^{(4)}_{\mathrm {diff},\,B} =\frac{4 a^2\epsilon}{a H}\R\R'(\partial\R)^2, \qquad \Ham^{(4)}_{\mathrm {diff},\,C}=\frac{2a^2\epsilon}{a H}\R^2\partial\R\partial\R'.
\end{align}
where the first term has been ``diff. induced" by the first operator in the Lagrangian \eqref{thirdlag}, while the other two terms have been ``diff. induced" by the last operator in Eq. \eqref{thirdlag}.
As will be discussed in Secs. \ref{sec:diffinduced2}
 and \ref{sec:explicitoneloop}, the addition of these quartic differential-induced interactions is essential to ensure that correlators with spatial derivatives, which do not satisfy consistency relations per se, combine within the loop integral to satisfy some consistency relations.

\subsection{Quadratic Hamiltonian from tadpoles}

From the Lagrangian in Eq. \eqref{thirdlag} one can derive the quadratic induced Hamiltonian from the linear tadpole counterterms---see discussion on section \eqref{sec:tad}\footnote{As discussed in Sec. \ref{sec:tad}, for the long-wavelength approximation discussed in this paper, we are only interested in terms where the operator outside the bracket in Eq. \eqref{tadpoleLag}: $\delta \Lt/\delta \R' \propto \R \R'$, i.e. from the second term in parenthesis in Eq. \eqref{thirdlag}. In the loop, $\R'$ would annihilates through the commutator with an external leg, while leaving a $\R$ with no derivative in the three-point function.}:
\begin{align}\label{tadpoleLag}
\Ham^{(2)}_1 &= -\frac{1}{2 a^2 \epsilon}\laa \frac{\delta \Lt}{\delta \R'}\raa \frac{\delta \Lt}{\delta \R'}=  -18\epsilon a^2\langle \R'\R \rangle \R'\R +\frac{9\epsilon a^2}{a H}\langle \R'^2 \rangle \R'\R+ \frac{6\epsilon a^2}{a H}\langle \R \, \partial^2 \R\rangle \R'\R.
\end{align}

\section{Consistency relations in non-slow-roll dynamics: explicit proofs}\label{sec:consistency}
We derive here a series of consistency relations valid for transient non-slow-roll phases of constant $\eta$. We consider long-wavelength modes that left the horizon well before the non-slow-roll phase and short modes which are not necessarily evaluated outside the horizon. The final goal would be to recognize these consistency relations inside the one-loop diagrams computed in the next section. 

To be able to perform a full analytical treatment, we consider, for definiteness, the case of instantaneous transitions between the slow-roll phases (where $\epsilon, |\eta| \ll 1$) and the non-slow-roll one in between. That is, we assume the following time dependent profile for the second slow-roll parameter:\footnote{Everything equally applies to the more generic case of asymmetric profiles, i.e. $\eta = \Delta\eta(\tau_s)(1-\theta(\tau_1-\tau_s))+  \Delta\eta(\tau_e)(1-\theta(\tau_1-\tau_e))
$.}
\be\label{eta}
\eta(\tau) = \Delta\eta\left[\theta(\tau-\tau_s)-\theta(\tau-\tau_e)\right]
\ee
where $\Delta\eta$ is a constant and $\tau_s$ and $\tau_e$ denote the beginning and the end of the non-slow-roll phase respectively.
We will highlight which part of the consistency relations depends on the sharpness of the transition and which one is generic to any phase of constant $\eta$.

We start by writing the generic solution for the mode functions in terms of Hankel functions.
To prove explicitly each consistency relation, we compute separetely the three-point correlator and the momentum derivative of the two-point function associated.
The reader not interested in these technical derivations may jump to the summary provided in Sec. \ref{summary} and proceed.

\subsection{Mode functions and Hankel functions relations}
 
The linear solution for $\R_k$ during a phase of constant-roll, i.e. constant $\nu = 3/2 +\eta/2$ (or generalizations), can be written as
\be
\R_k (\tau) = C_j x^{\nu_j}H^1_{\nu_j}(x)+D_j x^{\nu_j}H^2_{\nu_j}(x),
\ee
where $x = -k\tau$, $H_\nu^i$ ($i=1,2$) are the Hankel functions of the first and second type, while $C_j$ and $D_j$ are $k$-dependent coefficients. Following a transition from two constant-roll phases, these coefficients can be derived by imposing matching conditions to the field and its first derivatives at the transition point (see for instance \cite{Inomata:2021tpx,Fumagalli:2023loc}).
Assuming a change occurs at $\bar{x}_j = -k \tau_j$ from a phase with constant $\nu_{j-1}$ to one with constant $\nu_j$, this leads to
\renewcommand{\arraystretch}{1.1} 
\be\label{matrix0}
\left( \begin{array}{c}
C_j \\ D_j
 \end{array}
\right) = \mathbf{\mathcal{M}}_{\nu_{j-1}\nu_{j}}(\bar{x}_j) \left( \begin{array}{c}
C_{j-1} \\ D_{j-1}
 \end{array}
\right),
\ee
where we found convenient, for our purposes, to write the transition matrix as
\be\label{matrix1}
\mathbf{\mathcal{M}_{\nu_{j-1}\nu_{j}}}(\bar{x}_j)=\bar{x}_j^{\nu_1-\nu_2}\frac{i\pi\bar{x}_j}{4}\left( \begin{array}{cc}
    \{ H^1_{\nu_1} , H^2_{\nu_2} \} & \{ H^2_{\nu_1} , H^2_{\nu_2} \} \\
    -\{ H^1_{\nu_1} , H^1_{\nu_2} \} & -\{ H^2_{\nu_1} , H^1_{\nu_2} \}
\end{array} \right)\Bigg|_{\bar{x}_j},
\ee
and we have defined
\be\label{commHankel}
\{ H^i_{\alpha} , H^j_{\beta}\} \equiv  H^i_{\alpha} \cdot H^j_{\beta-1} - H^i_{\alpha-1} \cdot H^j_{\beta}.
\ee
The Bunch-Davies initial conditions in the infinite past set $D_1 = 0$ and 
\be
C_1 = \frac{H}{\sqrt{2 k^3}} \frac{i}{\sqrt{2 \epsilon_i  }}\sqrt{\frac{\pi}{2}}.
\ee
We thus
write, for instance, the mode functions after a transition from an initial slow-roll phase ($\nu\simeq 3/2$) to a generic constant-roll phase ($\nu$) as  

\be\label{modef}
\R_k(\tau) = C_1 \bar{x}^{\frac{3}{2}-\nu}   \frac{i\pi \bar{x}}{4} x^\nu\left( 
\acof_1(\bar{x})\cdot H_\nu^1(x)- \acof_2(\bar{x})\cdot H_\nu^2(x) \right),
\ee
with matching coefficients given by:
\be
\acof_1 (\bar{x})= \{ H^1_{3/2}(\bar{x}) , H^2_{\nu}(\bar{x})\},\quad \acof_2 (\bar{x}) = \{ H^1_{3/2}(\bar{x}) , H^1_{\nu}(\bar{x})\},
\ee 
and
\be\label{defx}
x = -k\tau,\quad \bar{x}= -k\tau_s,
\ee
with $\tau_s$ the time at which the generic constant-roll phase starts. Analogously, after the end of the non-slow-roll phase, for $\tau>\tau_e$, the mode functions are given by iterating \eqref{matrix0}-\eqref{matrix1}.
It is also beneficial to introduce the following relations satisfied by the Hankel functions:
\begin{align}\label{Hankelrel}
H^i_{\nu+1}(x) + H^i_{\nu-1}(x) = \frac{2 \nu}{x} H^i_{\nu}(x),\qquad
\frac{d}{dx}(x^{\nu}H^i_{\nu}(x) ) =x^{\nu} H^i_{\nu-1}(x).
\end{align}
Combining the two previous equations leads to
\be\label{derHankel}
\frac{d H^i_{\nu}}{dx} = \frac{ \nu}{x} H^i_{\nu}(x) -H^i_{\nu+1}(x),
\ee
and using Eq. \eqref{derHankel} and Eq. \eqref{Hankelrel} one can derive the following useful expression:
\be\label{Hankelrel2}
\frac{d}{dx}\{ H^i_{\alpha}(x), H^j_{\beta}(x)  \} = - \frac{\{ H^i_{\alpha}(x), H^j_{\beta}(x)  \}}{x} + \frac{\beta-\alpha}{x}(   H^i_{\alpha}(x), H^j_{\beta}(x) ),
\ee
where we have defined
\be \label{anticHankel}
(   H^i_{\alpha}, H^j_{\beta}      )  \equiv  H^i_{\alpha} \cdot H^j_{\beta-1} + H^i_{\alpha-1} \cdot H^j_{\beta}.
\ee

\subsection{Consistency relation for the bispectrum}

Maldacena consistency relation \cite{Maldacena:2002vr,Creminelli:2004yq,Cheung:2007sv} states that in the so called squeezed limit 
\be
p \ll k\simeq k',
\ee
the three-point functions satisfy the following relation
\begin{align}\label{Maldacena1}
\llangle  \Rh_{\bk}(\tau) \Rh_{\bk'}(\tau)\Rh_{\bp}(\tau) \rrangle &= -\frac{d\ln \p_\R (k,\tau)} {d\ln k} \frac{2\pi^2}{k^3}\p_\R(k,\tau)\frac{2\pi^2}{p^3}\p_\R(p,\tau)\\[1mm] \label{Maldacena2}
&= \left(-|\R_p(\tau)|^2\frac{2\pi^2}{k^3}\right)\frac{d\p_\R(k,\tau)} {d\ln k},
\end{align}
where we recall the notation:
\be
\langle \Rh_{\bp} \Rh_{\bk} \Rh_{\bk'} \rangle = (2\pi)^3\delta(\bp+\bk+\bk' )\llangle \Rh_{\bp} \Rh_{\bk} \Rh_{\bk'} \rrangle,
\ee
and
$\p_\R(k,\tau)$ is the dimensionless power spectrum, that we also rewrite here for practical purposes:
\be
\p_\R(k,\tau) = \frac{k^3}{2\pi^2}|\R_k(\tau)|^2.
\ee

\subsubsection{Squeezed bispectrum}

Let us first compute the left hand side of Eq. \eqref{Maldacena1}, i.e. the three-point function in the squeezed limit:
\begin{equation}\label{bispectrum}
\langle\hat{\R}_{\bk}(\tau) \hat{\R}_{\bk'}(\tau)\hat{\R}_{\bp}(\tau)\rangle=i\int_{\tauf}^{\tau}d\tau_{1}\langle [\Ha^{(3)}(\tau_{1}),\hat{\R}_{\bk}(\tau) \hat{\R}_{\bk'}(\tau)\hat{\R}_{\bp}(\tau)]\rangle,
\end{equation}
using the time-dependent profile for $\eta$ in Eq. \eqref{eta}.
We start by considering correlators among fields evaluated during the non-slow-roll phase ($\tau\in [\tau_s,\tau_e]$).
The bispectrum is more easily computed by using the rewriting of the Hamiltonian in Eq. \eqref{firstset}. With that, the $\eta'$ in the bulk term and the total derivative in the boundary terms allows one to easily integrate over time. Even so, we point out that one can use the equivalent form of the Hamiltonian in Eq. \eqref{secondset} and obtain the same results after integrating by parts.
The three interactions in Eq. \eqref{firstset} give respectively 
\begin{align}\label{bispectruma}
\llangle  \Rh_{\bk} \Rh_{\bk'} \Rh_{\bp}\rrangle_{a} &= -|\R_p(\tau)|^2 4 a^2(\tau_s)\epsilon(\tau_s)\cdot  \Delta\eta\cdot\Ima \left( \R^2_k(\tau) \R_k^*(\tau_s) \R_k'^*(\tau_s)\right) \equiv \mathcal{F}(\tau,\tau_s)\\[1mm] \label{bispectrumb} 
\llangle  \R_{\bk} \R_{\bk'}\R_{\bp} \rrangle_{b} &= |\R_p(\tau)|^2  |\R_k(\tau)|^2\cdot   \Delta\eta\\[1mm] \label{bispectrume}
\llangle  \R_{\bk} \R_{\bk'} \R_{\bp}\rrangle_{e} &= |\R_p(\tau)|^2 \frac{2 \Rea(\R_k(\tau)\R_k'^*(\tau))}{aH}.
\end{align}
Summing the three contributions provide a result in agreement with the findings obtained in \cite{Motohashi:2023syh}.
Note that the $\Delta\eta$ appearing in the first term corresponds to the size of the jump at the transition $\tau_s$, while the one on the second term is the constant value of $\eta$ during the non-slow-roll phase. Thus, considering the correlators for \(\tau > \tau_e\) results in adding \( -\mathcal{F}(\tau, \tau_e) \theta(\tau - \tau_e) \) to Eq. \eqref{bispectruma}, and substituting the time-dependent profile of \(\eta(\tau)\) from Eq. \eqref{eta} (which in this case is equivalent to setting \(\Delta\eta = 0\)) into Eq. \eqref{bispectrumb}.

Finally, note that last term is not slow-roll suppressed and in standard slow-roll cases (small $\eta$) would provide the dominant contribution to the bispectrum when short modes are on sub-horizon scale.

\subsubsection{Proof of the consistency relation}\label{proofbispectrum1}

The power spectrum during the non-slow-roll phase, i.e. $\tau \in [\tau_s,\tau_e]$, can be conveniently written, from Eq. \eqref{modef}, as 
\be  
\label{rewrpower}
\p_\R(k,\tau) = \frac{k^3}{2\pi^2}|\R_k(\tau)|^2 = \bar{x}^{2(3/2-\nu)} \cdot f(x,\bar{x}),
\ee
where
\be
f(x,\bar{x}) = \frac{k^3 |C_1|^2}{2\pi^2} \frac{\pi^2}{4^2}\cdot \bar{x}^2\cdot x^{2\nu} |\acof_1(\bar{x})\cdot H_\nu^1(x)- \acof_2(\bar{x})\cdot H_\nu^2(x) |^2,
\ee
and we remind that the combination $k^3|C_1|^2$ is $k$-independent.

Let us now compute the right-hand side of Eq. \eqref{Maldacena1} by taking the derivative of the power spectrum with respect to its momentum. We use

\be\label{dersplit}
\frac{d}{d\ln k} = x\frac{\partial}{\partial x} + \bar{x}\frac{\partial}{\partial\bar{x}} 
\ee
where $x,\bar{x}$ have been defined in Eq. \eqref{defx}. 
From Eq. \eqref{rewrpower}, we write

\be 
\label{splitofp}
\frac{d\p_\R(k,\tau)}{d\ln k} =  \bar{x}^{2(3/2-\nu)}\cdot \bar{x}\frac{\partial f(x,\bar{x})}{\partial \bar{x}}+   f(x,\bar{x})\cdot \bar{x}\frac{\partial \bar{x}^{2(3/2-\nu)}}{\partial \bar{x}} + \bar{x}^{2(3/2-\nu)}\cdot x\frac{\partial f(x,\bar{x})}{\partial x}.
\ee
We are now going to show that the three terms appearing on the right hand side, once inserted in Eq. \eqref{Maldacena1}, leads to the three contributions to the bispectrum in Eqs. \eqref{bispectruma}-\eqref{bispectrumb}-\eqref{bispectrume}. 
First, by reminding that $2(3/2-\nu) = -  \Delta\eta$, it is immediate to show that the second term leads to the contribution in Eq. \eqref{bispectrumb}, i.e. the one coming from the first boundary term in Eq. \eqref{firstset}: 
\be
-\frac{2\pi^2}{k^3}|\R_p(\tau)| ^2 f(x,\bar{x})\cdot \bar{x}\frac{\partial \bar{x}^{2(3/2-\nu)}}{\partial \bar{x}} =  |\R_p(\tau)|^2  |\R_k(\tau)|^2\cdot \Delta\eta =\llangle \Rh_{\bp} \Rh_{\bk} \Rh_{\bk'} \rrangle_{b}.
\ee
Note that, similar to Eq. \eqref{bispectrumb}, this term carries information solely on the constant value of $\nu$ (or $\eta$) at the time the power spectrum is evaluated, here $\tau \in [\tau_s,\tau_e]$.\footnote{When considering multiple transitions, starting from the Bunch-Davies initial condition ($\nu \simeq 3/2$) and evolving to a final value $\nu$, Eqs. \eqref{matrix0}-\eqref{matrix1} yield
$
\p_\R \propto (\bar{x}_1\bar{x}_2..\bar{x})^2 \bar{x}_1^{2(3/2 - \nu_1)} \bar{x}_2^{2(\nu_1 - \nu_2)} \cdots \bar{x}^{2(\nu_{j-1} - \nu)}.
$
The generalization of Eqs. \eqref{dersplit} then lead, for the second term in \eqref{splitofp}, to
$
2(3/2 - \nu_1 + \nu_1 - \nu_2 + \nu_2 \cdots + \nu) = -\eta,
$
where $\eta$ is the constant value of the second slow-roll parameter at the time the power spectrum is evaluated.
}

Second, we consider the last term in Eq. \eqref{splitofp} which is the one bringing memory of the time dependence of the power spectrum (i.e. $x=-k\tau$). In order to compare it with the contribution in Eq. \eqref{bispectrume}, we first write
\begin{align}\nonumber
2\cdot\Rea(\R_k(\tau)\R'^*_k(\tau)) = & |C_1|^2 \frac{\pi^2}{4^2}\cdot \bar{x}^{2(3/2-\nu)}\cdot \bar{x}^2\cdot x^{2\nu} \Big( (   H^1_{\nu}, H^2_{\nu}      )  (|\acof_1(\bar{x})|^2 + |\acof_2(\bar{x})|^2)\\[1mm] \label{rez} &- 4\mathrm{Re}\left( H^1_{\nu}(x)H^1_{\nu-1}(x)  \acof_1(\bar{x})\acof^*_2(\bar{x}) \right) \Big)\cdot(-k),
\end{align}
where we used
\be
\R'_k(\tau) = -k \partial_x \R_k \supset  -k \partial_x (x^\nu H_{\nu}(x)) =  -k x^\nu H_{\nu-1}(x).
\ee
Provided with Eq.\eqref{rez}, one can show that
\be
-\frac{2\pi^2}{k^3}|\R_p(\tau)| ^2 \bar{x}^{2(3/2-\nu)}\cdot x\frac{\partial f(x,\bar{x})}{\partial x} = -\tau\cdot 2\Rea(\R_k(\tau)\R'^*_k(\tau)) |\R_p(\tau)|^2 = \llangle \R_{\bp} \R_{\bk} \R_{\bk'} \rrangle_{e}, 
\ee
since $\tau \simeq -1/aH$.

Finally, a bit more work is required to prove that the first term in Eq. \eqref{splitofp} leads to the bispectrum contribution in Eq. \eqref{bispectruma}, i.e. the one coming from the bulk operator $\propto \eta' \R'\R^2$. 
In particular, we would like to show that the following equality holds
\begin{align}\label{partialforlast}
-\frac{2\pi^2}{k^3}|\R_p(\tau)|^2\bar{x}^{2(3/2-\nu)}\cdot \bar{x}\frac{\partial f(x,\bar{x})}{\partial \bar{x}} &=-|\R_p(\tau)|^24 a^2(\tau_s)\epsilon(\tau_s)\cdot  \Delta\eta\cdot\Ima \left( \R^2_k(\tau) \R_k^*(\tau_s) \R_k'^*(\tau_s)\right) 
\end{align}
By using the expression for the mode functions in Eq. \eqref{modef}, the right-hand side can be written as
\begin{align}
-4 a^2(\tau_s)\epsilon(\tau_s)\cdot  \Delta\eta\cdot\Ima \left( \R^2_k(\tau) \R_k^*(\tau_s) \R_k'^*(\tau_s)\right) |\R_p(\tau)|^2 = \mathcal{I}_1 \cdot  \mathcal{I}_2 |\R_p(\tau)|^2,
\end{align}
where
\be\label{I1}
\mathcal{I}_1 \equiv a^2(\tau_s)\epsilon(\tau_s) k|C_1|^4 \frac{\pi}{2} x^{2\nu }\bar{x}^{2(3-\nu)}
\ee
and
\be\label{I2}
\mathcal{I}_2 \equiv - 4\bar{x} \cdot  \Delta\eta\, \left( |H^1_{3/2}(\bar{x})|^2 \Ima^2\left(  H_\nu^1(x)  H_{\nu-1}^2(\bar{x})\right) - |H^1_{3/2-1}(\bar{x})|^2 \Ima^2\left(  H_\nu^1(x)  H_{\nu}^2(\bar{x})\right) \right). 
\ee
In addition, the left-hand side of Eq. \eqref{partialforlast} can be rearranged such as 
\be\label{part2}
\left(\frac{\pi^2}{4^2}|C_1|^2 x^{2\nu }\bar{x}\cdot \bar{x}^{2(3/2-\nu)}\right)\cdot\frac{\partial}{\partial \bar{x}} \left(-\bar{x}^2|\acof_1(\bar{x})\cdot H_\nu^1(x)- \acof_2(\bar{x})\cdot H_\nu^2(x) |^2\right) |\R_p(\tau)|^2.
\ee
Using the explicit expression for $|C_1|^2$, $\epsilon_i = \epsilon(\tau_s)$ and further manipulation in $\mathcal{I}_1$, one finds that the first term in parenthesis in the previous equation is equal to $\mathcal{I}_1$ as defined in Eq. \eqref{I1}.
To expand the derivative in the second term of Eq. \eqref{part2}, we use the Hankel function relations derived in Eq. \eqref{Hankelrel2} that we re-write here in terms of the mode functions coefficients:
\be\label{relcoefficient}
\frac{d}{d\bar{x}}\acof_i = - \frac{1}{\bar{x}}\acof_i + \left(\nu - \frac{3}{2}\right)\frac{1}{\bar{x}}\bar{\acof}_i,\quad i=1,2 ,
\ee
where we defined
\be
\bar{\acof}_1(\bar{x}) = (   H^1_{3/2}, H^2_{\nu}      ),\qquad \bar{\acof}_2(\bar{x}) = (   H^1_{3/2}, H^1_{\nu}      ).
\ee
Armed with the previous relations one finds
\begin{align}\label{derpar}
\frac{\partial}{\partial \bar{x}} &\left(-\bar{x}^2 |\acof_1(\bar{x})\cdot H_\nu^1(x)- \acof_2(\bar{x})\cdot H_\nu^2(x)|^2 \right) \nonumber
\\ &= \bar{x}\cdot  \Delta\eta\,  \Rea\Big( [H^1_\nu(x)]^2 (\bar{\acof}_1\acof_2^*+\bar{\acof}_2^*\acof_1) - H^1_\nu(x) H^2_\nu(x)(\bar{\acof}_1\acof_1^*+\bar{\acof}_2\acof_2^*) \Big).
\end{align}
Ultimately, by using the explicit expression for $\acof_i$ and $\bar{\acof}_i$ in terms of the Hankel functions, and after tedious algebraic manipulations, we obtain that the previous expression is equal to $\mathcal{I}_2$ as defined in Eq. \eqref{I2}. This concludes the proof of the equality \eqref{partialforlast}. 
We thus have proved explicitly that for a transition from slow-roll to a generic non-slow roll phase, the consistency relations is satisfied by means of the three operators singled out in the Hamiltonian \eqref{firstset}, i.e.
\be\label{finalb}
 \left(-|\R_p(\tau)|^2\frac{2\pi^2}{k^3}\right)\frac{d\p_\R(k,\tau)} {d\ln k} = \llangle \Rh_{\bp} \Rh_{\bk} \Rh_{\bk'} \rrangle_{a}+ \llangle \Rh_{\bp} \Rh_{\bk} \Rh_{\bk'} \rrangle_{b}+ \llangle \R_{\bp} \Rh_{\bk} \Rh_{\bk'} \rrangle_{e} = \llangle \Rh_{\bp} \Rh_{\bk} \Rh_{\bk'} \rrangle.
\ee
The generalization to transitions between two arbitrary phases is now trivial upon noticing that the term in parenthesis in Eq. \eqref{relcoefficient} is the difference between the two Hankel indices before and after the transition. 
Thus, the same consistency relation is derived when the correlators are evaluated after the non-slow-roll phase, by using the mode functions for \(\tau > \tau_e\) as obtained from Eqs. \eqref{matrix0}--\eqref{matrix1}, along with the updated version of the three-point functions, as detailed below Eqs. \eqref{bispectruma}--\eqref{bispectrumb}--\eqref{bispectrume}.

\subsection{Consistency relation for three-point functions with time derivatives}\label{derivativeconsistency} 
We would like to show that the three-point functions $\langle\R'_k\R'_k \R_p\rangle$ satisfies the following consistency conditions
\begin{align}\label{Maldacena3}
\llangle \Rh'_{\bk} (\tau)\Rh'_{\bk'} (\tau) \Rh_{\bp}(\tau) \rrangle &= -\frac{d\ln \p_{\R'} (k,\tau)} {d\ln k} \frac{2\pi^2}{k^3}\p_{\R'}(k,\tau)\frac{2\pi^2}{p^3}\p_\R(p,\tau)\\[1mm] 
&= \left(-|\R_p(\tau)|^2\frac{2\pi^2}{k^3}\right)\frac{d\p_{\R'}(k,\tau)} {d\ln k} 
\end{align}
where
\be
\p_{\R'}(k,\tau) = \frac{k^3}{2\pi^2}|\R'_p(\tau)|^2.
\ee

\subsubsection{Squeezed three-point function with time derivatives }\label{squeezedderivativeoperator} 
We start by computing the three-point function. As discussed in Sec. \ref{timederandquartic}, the key point to remember is that this correlator is not simply given by the trivial generalization of the commutator formula, but instead by Eq.
\eqref{correlatortwoderivative}:
\begin{align}\nonumber
\langle \Rh'_{\bk} \Rh'_{\bk'}  \Rh_{\bp} \rangle =&
i\int_{\tauf}^{\tau}d\tau_{1}\langle [\Ha^{(3)}(\tau_{1}),\hat{\R}'_{\bk}(\tau) \hat{\R}'_{\bk'}(\tau)\hat{\R}_{\bp}(\tau)]\rangle \\[1mm] \label{bispline2} &+ i \langle [\Ha^{(3)}(\tau),\hat{\R}_{\bk}(\tau)] \hat{\R}'_{\bk}(\tau)\hat{\R}_{\bp}(\tau) \rangle +  i \langle \hat{\R}'_{\bk}(\tau) [\Ha^{(3)}(\tau),\hat{\R}_{\bk}(\tau)] \hat{\R}_{\bp}(\tau) \rangle.
\end{align}
We compute the first line by means of the Hamiltonian expressed in the form given by Eq. \eqref{firstset}, where total derivative interaction terms allows one to easily integrate over time. However, to compute the second line, where no time integration has to be performed, the form \eqref{firstset} becomes rather inconvenient. We thus rely on the re-unfolded equivalent Hamiltonian simply derived from the Lagrangian in Eq. \eqref{thirdlag} where we only need to consider operators with time derivatives acting on $\R$. This is because the commutators on the second line are evaluated at equal times.

When computing the first line of Eq. \eqref{bispline2}, we use the same notation as in the previous section to label partial results. Specifically, subscripts indicate the part of the interaction Hamiltonian inserted in the first line of Eq. \eqref{firstset}: 
\be
\langle \Rh'_{\bk} \Rh'_{\bk'}  \Rh_{\bp} \rangle_j = 
i\int_{\tauf}^{\tau}d\tau_{1}\langle [\Ha_i^{(3)}(\tau_{1}),\hat{\R}'_{\bk}(\tau) \hat{\R}'_{\bk'}(\tau)\hat{\R}_{\bp}(\tau)]\rangle,\quad j = a,b,e
\ee
Following similar steps as in the previous section, the three operators in Eq. \eqref{firstset} gives respectively, for $\tau \in [\tau_s,\tau_e]$:
\begin{align}\label{bispectrumdera}
\llangle  \Rh'_{\bk} \Rh'_{\bk'} \Rh_{\bp} \rrangle_{a} &= -|\R_p(\tau)|^2 4 a^2(\tau_s)\epsilon(\tau_s)\cdot  \Delta\eta\cdot\Ima \left( \R'^2_k(\tau) \R_k'^*(\tau_s) \R_k^*(\tau_s)\right)\equiv \widetilde{\mathcal{F}}(\tau,\tau_s) \\[1mm] \label{bispectrumderb}
\llangle  \Rh'_{\bk} \Rh'_{\bk'} \Rh_{\bp}\rrangle_{b} &= -|\R_p(\tau)|^2  |\R'_k(\tau)|^2\cdot   \Delta\eta\\[1mm] \label{bispectrumdere}
\llangle  \Rh'_{\bk} \Rh'_{\bk'} \Rh_{\bp}\rrangle_{e} &\propto |\R'_p(\tau)|\simeq 0.
\end{align}
Eq. \eqref{bispectrumdera} is what one would have naively guessed from \eqref{bispectruma}. Note that there is no inconsistency between the minus sign difference in Eq. \eqref{bispectrumderb} and in Eq. \eqref{bispectrumb}. After integrating the total derivative term interaction, the non-zero contributions are proportional to the equal-time commutator $[\R_k,\R'_k]$ in the former case and $ [\R'_k,\R_k]$ in the latter. 
For the same reason, the contribution in Eq. \eqref{bispectrumdere} is proportional to $|\R'_p(\tau)|^2$ instead of $|\R_p(\tau)|^2$ and therefore negligible in the squeezed limit ($p$ is the long-wavelength mode that is well super-horizon at the time $\tau$).

Let us now turn to the contact terms, i.e. the second line in Eq. \eqref{bispline2}. For this purpose, we use the cubic Hamiltonian derived from the Lagrangian in Eq. \eqref{thirdlag}, specifically focusing on the terms containing time derivatives, which we re-write explicitly here:
\be\label{Heff}
\Ha^{(3)}_{\rm eff} \equiv\int d^3 x \left( \frac{\epsilon a^2}{aH}\R'^3  -3\epsilon a^2\R'^2\R + \frac{2 a^2\epsilon}{a H}\R'\R\partial^2 \R \right).
\ee
Inserting $\Ha_{\rm eff}$ in the second line of Eq. \eqref{bispline2}, we arrive at
\begin{align}\nonumber
i \llangle &[\Ha_{\rm eff}^{(3)}(\tau),\hat{\R}_{\bk'}(\tau)] \hat{\R}'_{\bk}(\tau)\hat{\R}_{\bp}(\tau) \rrangle +  i \llangle \hat{\R}'_{\bk'}(\tau) [\Ha_{\rm eff}^{(3)}(\tau),\hat{\R}_{\bk}(\tau)] \hat{\R}_{\bp}(\tau) \rrangle   \\[1mm] \label{contact1}
&= -6 |\R_p(\tau)|^2|\R'_k(\tau)|^2 -\frac{2 k^2}{a H} |\R_p(\tau)|^2 \Rea(\R_k(\tau)\R'^*_k(\tau))
\end{align}
The first term\footnote{The quartic induced Hamiltonian coming from this piece is $\propto \R'^4$ and analogously gives suppressed contribution to the one-loop power spectrum of the small momentum $p$.} in Eq. \eqref{Heff} gives a negligible contribution since it is $\propto |\R'_p(\tau)|^2$, while the first and second term in Eq. \eqref{contact1} are derived respectively from the second and third terms in $\Ha_{\rm eff}$ in Eq. \eqref{Heff}.
This contact interaction contributions are the one that in the loop diagrams appear thanks to the quartic induced Hamiltonian. Without including the latter, one would not be able to write the one-loop correction as the integral of a three-point function and then use the consistency relation.
Putting all together, we derive the final expression:
\begin{align}\label{bispectrumder0}
\llangle \R'_{\bk} \R'_{\bk'} \R_{\bp}\rrangle  =& -|\R_p(\tau)|^2 4 a^2(\tau_s)\epsilon(\tau_s)\cdot  \Delta\eta\cdot\Ima \left( \R'^2_k(\tau) \R_k'^*(\tau_s) \R_k^*(\tau_s)\right)\\[1mm] \label{bispectrumderfull}
&- (  \Delta\eta+6)|\R_p(\tau)|^2  |\R'_k(\tau)|^2 -\frac{2 k^2}{a H} |\R_p(\tau)|^2 \Rea(\R_k(\tau)\R'^*_k(\tau)).
\end{align}
As before, $\Delta\eta$ in the first term corresponds to the size of the jump at the transition $\tau_s$, while the one on the second line, coming from the contribution in Eq. \eqref{bispectrumderb}, is the constant value of $\eta$ during the non-slow-roll phase. Thus, the above correlator after the non-slow-roll phase, i.e. $\tau>\tau_e $, is simply given by adding 
$-\widetilde{\mathcal{F}}(\tau_e,\tau_1)\theta(\tau_1-\tau_e)$ (see definition in Eq. \eqref{bispectrumdera}) and substituting $\eta(\tau)$ in Eq. \eqref{eta}  (sending $\Delta\eta=0$) into the second line.

\subsubsection{Proof of the consistency relation with time derivatives} 

In order to show that the three-point function in Eq. \eqref{bispectrumderfull} is equal to the right-hand side of Eq. \eqref{Maldacena3}, we follow similar steps as in the previous section.
We start by writing the dimensionless power spectrum of the field derivative during the non-slow-roll phase as
\be\label{pzetapr}
\p_{\R'}(k,\tau) = \frac{k^3}{2\pi^2}|\R'_k(\tau)|^2 = k^2\cdot \bar{x}^{2(3/2-\nu)} \cdot  f_{\R'}(x,\bar{x}),
\ee
where
\be
f_{\R'}(x,\bar{x}) = \frac{k^3 |C_1|^2}{2\pi^2} \frac{\pi^2}{4^2}\cdot \bar{x}^2\cdot x^{2\nu} |\acof_1(\bar{x})\cdot H_{\nu-1}^1(x)- \acof_2(\bar{x})\cdot H_{\nu-1}^2(x) |^2,
\ee
Using the splitting in Eq. \eqref{dersplit} (and taking into account the explicit $k$-dependence in Eq. \eqref{pzetapr}), gives us
\be 
\label{splitofpder}
\frac{d\p_{\R'}(k,\tau)}{d\ln k} =  \bar{x}^{2(3/2-\nu)}\cdot \bar{x}\frac{\partial f_{\R'}(x,\bar{x})}{\partial \bar{x}}+ (2 -  \Delta\eta) \p_{\R'} + \bar{x}^{2(3/2-\nu)}\cdot x\frac{\partial f_{\R'}(x,\bar{x})}{\partial x}.
\ee
The first term on the right-hand side leads to the bispectrum contribution on the first line in Eq. \eqref{bispectrumderfull} in complete analogy with the first term in \eqref{splitofp} leading to Eq. \eqref{bispectruma}. We can simply borrow the laborious proof of section \ref{proofbispectrum1} and replace everywhere $H_{\nu}(x)\rightarrow H_{\nu-1}(x)$.

The second term in \eqref{splitofpder} is simply derived by taking the derivative of the first two factors multiplying $f$ in \eqref{pzetapr}. It remains to be shown that the sum of this term and the last one results in the second line in Eq. \eqref{bispectrumderfull}.

First, we apply a straightforward manipulation to $f_{\R'}$:
\be
f_{\R'}(x,\bar{x}) \supset  x^{2\nu} \cdot H_{\nu-1}^i(x)  H_{\nu-1}^j(x) =  x^2 x^{2(\nu-1)} \cdot H_{\nu-1}^i(x)  H_{\nu-1}^j(x) 
\ee 
then use \eqref{Hankelrel} and $\eqref{derHankel}$ to obtain
\be
\frac{d}{dx}(x^{\nu-1}H^i_{\nu-1}(x) ) =x^{\nu-1} H^i_{\nu-2}(x) = x^{\nu-1}\left(-H_\nu +\frac{2(\nu-1)}{x}H_{\nu-1}\right).
\ee
Armed with the two previous equations, using $4(\nu-1)+2 = 4+2  \Delta\eta$ and reminding the expression in Eq. \eqref{rez} one may find 

\be
\bar{x}^{2(3/2-\nu)}\cdot x\frac{\partial f_{\R'}(x,\bar{x})}{\partial x} =  (4+2  \Delta\eta)\p_{\R'}
 +\frac{k^3}{2\pi^2}\frac{2 k^2}{a H}\Rea(\R_k(\tau)\R'^*_k(\tau)),
\ee
Putting all together we successfully arrived at
\be
-\frac{2\pi^2}{k^3}|\R_p(\tau)|^2 \frac{d\p_{\R'}(k,\tau)}{d\ln k} = \llangle \R'_{\bk} \R'_{\bk'} \R_{\bp} \rrangle,
\ee
where the right-hand side is provided in Eq. \eqref{bispectrumderfull}.
For the generalization to arbitrary transitions, the same comments as those below Eq. \eqref{finalb} apply.

\subsection{Consistency relations for operators with spatial derivatives }
\label{sec:diffinduced2}

Let us now review how a term from the cubic Hamiltonian, $\Hamt = -\Lt$, and its diff. induced quartic counterpart--see Section \ref{sec:diffinduced}, $\Ham^{(4)}_{\rm diff} =  - \Lag_{\rm diff}^{(4)}$, contribute to one-loop diagrams by taking, as an example, the operators in Eq. \eqref{firstdiffex}. 

Schematically one has
\begin{align} \nonumber
\langle \Rh_{\bp}\Rh_{\bp'}\rangle_{\Ha^{(3)}}\,=\, &-\int^{\tau}d\tau_{1}\int^{\tau_{1}}d\tau_{2}\int d\bK_1\langle [\Ha^{(3)}(\tau_{2}),[c_1 \Rh_{\bk_{1,1}}(\partial_i\Rh)_{\bk_{1,2}}(\partial_i\Rh)_{\bk_{1,3}},\hat{\R}_{\bp}\hat{\R}_{\bp'}]]\rangle\\[1mm] \label{diff1}
= \, & 2i \int^\tau d\tau_{1}\int d\bK_1 c_1  [\Rh_{\bk_{1,1}},\Rh_{\bp'}] \cdot\left( \langle 
 (\partial_i\Rh)_{\bk_{1,2}}(\partial_i\Rh)_{\bk_{1,3}}\Rh_{\bp} \rangle\right),
\end{align}
where $d\bK_1$ follows our usual notation in Eq. \eqref{bigK2}, and we leave here explicit the subscripts on the momenta.
From the quartic diff. induced operator one instead obtains 
\begin{align} \nonumber
\langle \Rh_{\bp}\Rh_{\bp'}\rangle_{\Ha^{(4)}_{\rm diff}}\,=\,&i\int^{\tau}d\tau_{1}\langle[\Ha_{\rm diff}^{(4)}(\tau_1),\hat{\R}_{\bp}\hat{\R}_{\bp'}]\rangle\\[1mm] \label{diff2}
=\,& 2i \int^\tau d\tau_{1}\int d\bK_1 c_1 [\Rh_{\bk_{1,1}},\Rh_{\bp'}]\cdot \left(-2 \langle 
 \Rh_{\bk_{1,1}}(\partial_i\Rh)_{\bk_{1,2}}(\partial_i\Rh)_{\bk_{1,3}} \Rh_{\bp}\rangle  \right)   ,
\end{align}
The two terms in parenthesis in Eqs \eqref{diff1} and \eqref{diff2} sum up and satisfy a consistency conditions. 
Let us write them explicitly: 
\be
\llangle  (\partial\R)_{\bk}(\partial\R)_{\bk'}\R_{\bp} \rrangle = +k^2 \llangle \R_{\bp} \R_{\bk} \R_{\bk'} \rrangle =  \left(-|\R_p(\tau)|^2\frac{2\pi^2}{k^3}\right)\cdot k^2\,\frac{d\p_\R(k,\tau)} {d\ln k} 
\ee
where we used the previously proven consistency relation in Eq. \eqref{Maldacena2},
and
\be
-2\llangle 
 (\partial_i\Rh)_{\bk'}(\partial_i\Rh)_{\bk} \Rh_{\bp}\Rh_{\bp'}\rrangle =    -2\,k^2 |\R_p(\tau)|^2|\R_k|^2 =  \left(-|\R_p(\tau)|^2\frac{2\pi^2}{k^3}\right)\cdot 2 k^2\,\p_\R(k,\tau).
\ee
By summing the two previous equations, one obtains the following consistency conditions
\be
\llangle  (\partial\R)_{\bk}(\partial\R)_{\bk'}\R_{\bp} \rrangle-2\llangle 
 (\partial_i\Rh)_{\bk'}(\partial_i\Rh)_{\bk} \Rh_{\bp}\Rh_{\bp'}\rrangle = \left(-|\R_p(\tau)|^2\frac{2\pi^2}{k^3}\right)\cdot \,\frac{d \left(\,k^2\p_\R(k,\tau)\right)} {d\ln k}.
\ee
By combining results of the previous sections one could analogously consider consistency relations for operator with time and spatial derivatives.

\subsection{Summary of consistency relations}\label{summary}
Let us summarize in an unified manner the consistency relations found in this section for correlators evaluated during or after the non-slow-roll phase. We label them for future reference as follows:
\begin{align}\nonumber
\mathcal{A}_0 &= |\R_p|^{-2}  \llangle\Rh_{\bk}(\tau_1)\Rh_{\bk'}(\tau_1)\Rh_{\bp}(\tau_1) \rrangle = -\frac{2\pi^2}{k^3}\cdot\frac{d\p_\R(k,\tau_1)}{d\ln k} \\[1mm] \label{A0}&= \eta |\R_k(\tau_1)|^2 +\frac{2\Rea(\R_k(\tau_1)\R'^*_k(\tau_1))}{a H}+ \mathcal{F}(\tau_s,\tau_1) - \mathcal{F}(\tau_e,\tau_1)\theta(\tau_1-\tau_e)\\[1.2mm] \label{A}
\mathcal{A} &= |\R_p|^{-2} \partial_{\tau_1}\llangle\Rh_{\bk}(\tau_1)\Rh_{\bk'}(\tau_1)\Rh_{\bp}(\tau_1) \rrangle 
\\[1mm] \nonumber &=\frac{2}{a H} |\R'_k(\tau_1)|^2 -6 \,\Rea(\R_k(\tau_1)\R'^*_k(\tau_1)) -\frac{2 k^2}{a H} |\R_k(\tau_1)|^2 \\[1mm]&\qquad+\partial_{\tau_1}\mathcal{F}(\tau_1,\tau_s)- \partial_{\tau_1}\mathcal{F}(\tau_1,\tau_e)\theta(\tau_1-\tau_e)\\[1.2mm]
\nonumber
\mathcal{B} &=|\R_p|^{-2} \llangle\Rh'_{\bk}(\tau_1)\Rh'_{\bk'}(\tau_1)\Rh_{\bp}(\tau_1) \rrangle = -\frac{2\pi^2}{k^3}\frac{d\p_{\R'}(k,\tau_1)}{d\ln k}\\[1mm] \label{B}
&= -(\eta + 6)|\R'_k(\tau_1)|^2 -\frac{2 k^2}{a H}\Rea(\R_k(\tau_1)\R'^*_k(\tau_1)) + \widetilde{\mathcal{F}}(\tau_1,\tau_s)- \widetilde{\mathcal{F}}(\tau_1,\tau_e)\theta(\tau_1-\tau_e)\\[1.2mm]
\label{C}
\mathcal{C} & = -\frac{2\pi^2}{k^3}\frac{d \left(k^2\p_{\R}(k,\tau_1)\right)}{d\ln k} = k^2\mathcal{A}_0 -2 k^2|\R_k(\tau_1)|^2\\[1.2mm]
\label{D}
\mathcal{D}&=  -\frac{2\pi^2}{k^3}\cdot\partial_{\tau_1}\left(\frac{ k^2 d\p_\R(k,\tau_1)}{d\ln k} \right) = k^2 \mathcal{A} -4 k^2\Rea(\R_k(\tau_1)\R'^*_k(\tau_1)),
\end{align}
where
\be\label{F1}
\mathcal{F}(\tau_1,\bar{\tau}) = -4 a^2(\bar{\tau})\epsilon(\bar{\tau})\cdot\Delta\eta\cdot\Ima \left( \R^2_k(\tau_1) \R_k^*(\bar{\tau}) \R_k'^*(\bar{\tau})\right),
\ee
and
\be\label{F2}
\widetilde{\mathcal{F}}(\tau_1,\bar{\tau})= - 4 a^2(\bar{\tau})\epsilon(\bar{\tau})\cdot\Delta\eta\cdot\Ima \left( \R'^2_k(\tau_1) \R_k^*(\bar{\tau}) \R_k'^*(\bar{\tau}) \right).
\ee
All background quantities in the equations above are evaluated at $\tau_1$ unless specified otherwise, also $\eta$ appearing in the previous expressions (note not the constant $\Delta\eta$) is the time-dependent profile given in Eq. \eqref{eta}, namely contributions proportional to this value are negligible for $\tau_1>\tau_e$. 
As discussed, expressions \eqref{C} and \eqref{D}, which are trivially derived from the ones just above, will appear inside loop diagrams when combining cubic and diff. induced quartic interactions.
Further, consistency relations for correlator with one derivative operator can be derived, as in Sec. \ref{derivativeconsistency}, by using Eq. \eqref{correlatoronederivative}. Equivalently, one can also simply take the derivative of \eqref{A0} and \eqref{C}, because in the squeezed approximation $\partial_{\tau_1}\llangle\R_k\R_k \R_p\rrangle \simeq \llangle\R'_k\R_k \R_p\rrangle + \llangle\R_k\R'_k \R_p\rrangle.
$
Using the eom on the mode functions one arrives at \eqref{A}-\eqref{D}.

Let us comment on how general these results are.
The functions $\mathcal{F}$ and $\widetilde{\mathcal{F}}$ are the only terms keeping memory of the sharpness of the transition, here assumed to be instanteneous to do explicit analytical computations. In particular, in computing the bispectrum these terms are obtained by taking $\eta'$ as a superposition of Dirac deltas, while in taking derivative of the power spectrum, these contributions come from the momentum derivative of the matching coefficients (which are obtained assuming an instantenous transition). 
All other pieces are generic for a phase of constant $\eta$. 

For most of the one-loop diagrams we will consider the consistency relations during the non-slow-roll phase are sufficient for our purpose. It was however trivial to generalize the previous relations when fields are evaluated at arbitrary times after the non-slow-roll phase, including $\tau_1>\tau_e$. This boiled down, in the expressions above, to add
\be
-\mathcal{F}(\tau_e,\tau_1)\theta(\tau_1-\tau_e),\qquad -\widetilde{\mathcal{F}}(\tau_e,\tau_1)\theta(\tau_1-\tau_e),
\ee
and promote $\eta$ to the time-dependent profile in Eq. \eqref{eta}.

\section{One-loop corrections and consistency relations in non-slow-roll dynamics}\label{sec:explicitoneloop}
As discussed in Sec. \ref{sec:origin scale invariant}, we are interested in relative scale-invariant one-loop contributions to the power spectrum of the form highlighted in Eqs. \eqref{splitdiagram2}-\eqref{scaleinvariant}. These are the would be dominant correction in the limit in which the external momentum $\bp$ is much smaller than the ones running in the loops $\bk$, and arise by considering interactions active at times where the long-wavelength associated with $p$ is well outside the horizon. These corrections take the form:
\be\label{scaleinvariant2}
\p_{\R}^{\mathrm{1-loop}}(p)=\frac{p^3}{2\pi^2}\llangle \Rh_{\bp}\Rh_{\bp'}\rrangle = \p^{\mathrm{tree}}(p)\int d\tau_1\int d k \, C(k,\tau_1).
\ee
In this section, we first provide results, reformulated into a convenient and compact form, from brute-force calculations of one-loop diagrams derived from the cubic and quartic Hamiltonians discussed in Sec. \ref{sec:Hamiltonian}. This encompasses the general case of a non-slow-roll phase with arbitrarily $\eta$ (and first order in $\epsilon$) and can easily be extended to other scenarios. We then explicitly show how combining these results lead to an expression involving a time integration of three-point functions plus additional terms which enable us to apply the consistency relations established in the previous section and summarized in Eqs. \eqref{A}-\eqref{B}-\eqref{C}-\eqref{D}.
As in the previous section, we use, when necessary, the instantaneous transitions profile for $\eta$ as given in Eq. \eqref{eta}.

\subsection{One-loop corrections from the cubic Hamiltonian}

We compute one-loop corrections to the power spectrum at late time $\tau$, using the Hamiltonian detailed in Sec. \ref{sec:Hamiltonian}, in the form given by Eqs. \eqref{secondset}. We will comment on how to pass explicitly from results obtained through the Hamiltonian in the form in Eq. \eqref{secondset} to the ones obtained through Eqs. \eqref{firstset}. This allows one to perform an integration over time (specifically, \(\tau_2\) in our labeling) without needing to specify the mode functions, thereby maintaining the generality of the treatment and facilitating combinations with the findings of the previous sections.
This part complements our previous work \cite{Fumagalli:2023hpa}, where only terms not proportional to the Green's function associated with the linear eom of $\R$, i.e. $g_p$, from the first two interaction operator in Eqs. \eqref{firstset} and \eqref{secondset},  were considered. 

We use the short-hand notation:
\be\label{notationcomm}
\langle \zeta_{\bp} \zeta_{\bp'}\rangle_{[i,j]}\equiv -\int^{\tau} d\tau_1\int^{\tau_1}d\tau_2 [H_i(\tau_2),[H_j(\tau_1),\zeta_{\bp}\zeta_{\bp'}]].
\ee
To express results in a convenient form, we utilize the equal-time and unequal-time commutator relations from Eqs. \eqref{commutator01}-\eqref{Wronskian}-\eqref{commutatorsgreen}. Specifically, we recall that \be
g_p(\tau,\tau_1) = 2i \Ima(\R_p(\tau_1)\R_p^*(\tau))\Wr^{-1},\ee where the Wronskian is given by $\Wr = i/(2 a^2 \epsilon)$.

The contributions from $\mathcal{H}_c^{(3)}$ and $\mathcal{H}_d^{(3)}$ in Eqs. \eqref{secondset}, in the limit $p \ll k$, can be written, after several manipulations\footnote{Once all the Wick contractions have been performed, one can integrate by parts one of the $\R'(\tau_2)$ present in $[c,c]$ and use the equation of motion. This leads to a term proportional to $\eta'$, a total derivative term (in $\tau_2$) and a term that cancels the contribution from $[c,d]$. We can thus perform the integral over $\tau_2$, also explicitly using Eq. \eqref{eta}. The two left terms are the one equivalently arising from a computation with the cubic Hamiltonian in Eq. \ref{firstset}.}, in the following compact form: 
\begin{align}\nonumber
\llangle\zeta_{\bp}  \zeta_{\bp'}\rrangle_{[c + d,c]  }&\equiv\llangle\zeta_{\bp}  \zeta_{\bp'}\rrangle_{[c ,c]  } +\llangle\zeta_{\bp}  \zeta_{\bp'}\rrangle_{[d,c]  }\\[1mm] \label{cd,c}
&=|\R_p(\tau)|^2   \int^{\tau}d\tau_1 \int \frac{d\bk}{(2\pi)^3}\eta\cdot\left(  g_p(\tau,\tau_1)\left( \,   \eta |\zeta'_{k}(\tau_1)|^2 -\widetilde{\mathcal{F}}(\tau_s,\tau_1)\right)+\partial_{\tau_1}\mathcal{F}(\tau_s,\tau_1)\right), 
 \end{align}
and
\begin{align}\label{cd,d}
\llangle \zeta_{\bp}\zeta_{\bp'}\rrangle_{[c + d,d]  }&=\, |\R_p(\tau)|^2   \int^{\tau}d\tau_1\ \int \frac{d\bk}{(2\pi)^3}\,\eta\cdot g_p(\tau,\tau_1)\left( \eta \cdot k^2  |\zeta_k(\tau_1)|^2  + k^2 \mathcal{F}(\tau_1,\tau_s)\right),
\end{align}
where $\mathcal{F}$ has been defined in Eq. \eqref{F1}.
Throughout the all section, the mode functions appearing inside the integrals are meant to be evaluated at $\tau_1$, while the contribution proportional to the tree-level power spectrum $|\R_p|^2$ is consider evaluated at the end of inflation $\tau$. This is simply because $p$ is well outside the horizon at the time interactions are active and $\mathrm{Re}\left(\zeta_p(\tau_1)\zeta^*_p(\tau)\right) \simeq |\R_p(\tau)|^2$.

Let us now turn to the contributions coming from the last boundary term within the set in Eqs.\eqref{secondset}. When this latter appear in the ``external" nested commutator, i.e. $\mathcal{H}^{(3)}(\tau_2) = \mathcal{H}^{(3)}_e $, one can simply integrate the total derivative term and obtain:
\begin{align}\label{e,c}
    \llangle\zeta_{\bp}  \zeta_{\bp'}\rrangle_{[e,c]} &=   |\R_p|^2   \int^{\tau}d\tau_1 \int \frac{d\bk}{(2\pi)^3} \left(  \eta\cdot \frac{2}{a H}|\R'_k|^2 \right),
\end{align}
and
\begin{align} \label{e,d}
    \llangle\zeta_{\bp}  \zeta_{\bp'}\rrangle_{[e,d]} &= |\R_p|^2   \int^{\tau}d\tau_1 \int \frac{d\bk}{(2\pi)^3} \left(  \eta\cdot \frac{2 k^2}{a H}\Rea(\R_k\R'^*_k)  g_p(\tau,\tau_1) \right).
\end{align}
A bit more work is required to finally obtain the contributions $\llangle\zeta_{\bp}  \zeta_{\bp'}\rrangle_{[i,e]}$ where $i = c,d,e$. There we re-expand the total derivative in $\mathcal{H}^{(3)}_e(\tau_1) $ and perform the computation for each term separately. After tedious manipulations, similar to the ones leading to \eqref{cd,c}-\eqref{cd,d}, we obtain:
\begin{align}\nonumber
    \llangle \zeta_{\bp}  \zeta_{\bp'}\rrangle_{[c+d, e]} &= |\R_p|^2   \int^{\tau} \int \frac{d\bk}{(2\pi)^3} \Big( -\eta(\eta+3) |\R'_k|^2 g_p(\tau,\tau_1) -\eta\cdot \frac{2 k^2}{a H} |\R_k|^2 -3\eta\cdot \frac{1}{a H}|\R'_k|^2 \\[1mm]\label{cd,e} &  + \mathcal{O}_1(\tau_1,\tau_s) + g_p(\tau,\tau_1)\mathcal{O}_2(\tau_1,\tau_s) - \theta(\tau_1-\tau_e)\left(\mathcal{O}_1(\tau_1,\tau_e) + g_p(\tau,\tau_1)\mathcal{O}_2(\tau_1,\tau_e)\right)\Big)
\end{align}
where
\begin{align}
\mathcal{O}_1(\tau_1,\tau')&=- (\eta+3)\partial_{\tau_1}\mathcal{F}(\tau_1,\tau') +\frac{3}{a H} \widetilde{\mathcal{F}}(\tau_1,\tau') - \frac{2 k^2}{a H} \mathcal{F}(\tau_1,\tau')  \\[1mm]
\mathcal{O}_2(\tau_1,\tau')&=
(\eta+3)\widetilde{\mathcal{F}}(\tau_1,\tau_s) + \frac{ k^2}{a H} \partial_{\tau_1}\mathcal{F}(\tau_1,\tau'),
\end{align}
and $\mathcal{F},\widetilde{\mathcal{F}}$ have been defined in Eqs. \eqref{F1},\eqref{F2}.
Note that the dependence on $\tau_e$, the time at which the non-slow-roll phase ends, has vanished in Eqs. \eqref{cd,c}-\eqref{cd,d} because $\mathcal{H}^{(3)}_c(\tau_1)$ and $\mathcal{H}^{(3)}_d(\tau_1)$ are both proportional to $\eta$ forcing the integral over time to be constrained within the non-slow-roll phase. Thus, diagrams $[e,c+d+e],$ are the only places where contributions depending on $\tau_e$ may explicitly appear. These terms, combined with the rest, ensures that the consistency relations is satisfied also when considering the integrands after $\tau_e$.

For the last cubic contributions we finally have:
\begin{align}\label{e,e}
    \llangle \zeta_{\bp}  \zeta_{\bp'}\rrangle_{[e, e]} =& |\R_p|^2   \int^{\tau} \int \frac{d\bk}{(2\pi)^3} \left( -\frac{4 k^2}{a^2 H^2}\Rea(\R_k\R'^*_k) - (\eta+3)\cdot\frac{2}{a H}|\R'_k|^2 \right.\\[1mm]
   &\left.  \qquad\qquad +\frac{2 k^2}{a^2 H^2}    |\R'_k|^2 g_p(\tau,\tau_1) \right).
\end{align}

\subsection{One-loop corrections from the quartic Hamiltonian}

We list below the various contributions to the one-loop power spectrum obtained from the quartic Hamiltonian detailed in Sec. \ref{sec:Hamiltonian}.
\subsection*{One-loop corrections from quartic induced Hamiltonian}

Building on the components of the Hamiltonian defined in Eqs. \eqref{quarticinducedset}-\eqref{quarticinducedset1}-\eqref{quarticinducedset2}, we adopt the following notation:
\begin{align}
\langle \Rh_{\bp}\Rh_{\bp'}\rangle_{\Ha^{(4)}_{3,X}}\,\equiv\,i\int^{\tau}d\tau_{1}\langle[\Ha^{(4)}_{3,X}(\tau_1),\hat{\R}_{\bp}\hat{\R}_{\bp'}]\rangle,
\end{align}
and obtain:
\begin{align} \label{q1}
\llangle \zeta_{\bp}  \zeta_{\bp'}\rrangle_{H^{(4)}_{3,A}}  &= |\R_p|^2  \int^{\tau}d\tau_1\int \frac{d\bk}{(2\pi)^3}\, 9\, \left( \textcolor{red}{4} \cdot\Rea(\R_k\R_k'^*) -2 |\R'_k|^2  g_p(\tau,\tau_1)\right),\\[1mm]
\llangle \zeta_{\bp}  \zeta_{\bp'}\rrangle_{H^{(4)}_{3,B}}  &= |\R_p|^2  \int^{\tau}d\tau_1\int\frac{d\bk}{(2\pi)^3} \, \left( -\frac{k^4}{a^2H^2}  |\R_k|^2 g_p(\tau,\tau_1)\right),\\[1mm] \label{q2}
\llangle \zeta_{\bp}  \zeta_{\bp'}\rrangle_{H^{(4)}_{3,C}}  &= |\R_p|^2  \int^{\tau}d\tau_1\int\frac{d\bk}{(2\pi)^3} \, 9\,\left( -\frac{\textcolor{red}{3}}{a H} |\R'_k|^2\right),\\[1mm] \label{q3}
\llangle \zeta_{\bp}  \zeta_{\bp'}\rrangle_{H^{(4)}_{3,D}}  &= |\R_p|^2  \int^{\tau}d\tau_1\int\frac{d\bk}{(2\pi)^3} \, 3\,\left( \textcolor{red}{4}\cdot\frac{k^2}{a H}|\R_k|^2 -4 \frac{k^2}{a H} \Rea(\R_k\R_k'^*)g_p(\tau,\tau_1) \right),\\[1mm]
\llangle \zeta_{\bp}  \zeta_{\bp'}\rrangle_{H^{(4)}_{3,E}}  &= |\R_p|^2  \int^{\tau}d\tau_1\int\frac{d\bk}{(2\pi)^3} \, 6\,\left(- \frac{k^2}{a^2H^2}\Rea(\R_k\R_k'^*) \right).
\end{align}
All mode functions inside the integral on the right-hand sides are evaluated at $\tau_1$.
\subsection*{One-loop corrections from quadratic tadpole induced Hamiltonian}
Contributions from the tadpole induced Hamiltonian in Eq. \eqref{tadpoleLag} --see Section \ref{sec:tad}, leads to the following one-loop contributions to the power spectrum:
\be
\llangle \zeta_{\bp}  \zeta_{\bp'}\rrangle_{H^{(2)}_{1}} = |\R_p|^2\int^{\tau}d\tau_1\int\frac{d\bk}{(2\pi)^3} \,  \left( -9\cdot2\,\Rea(\R_k\R_k'^*) + 9\cdot\frac{1}{a H} |\R'_k|^2 -3\cdot 2\frac{k^2}{a H } |\R_k|^2    \right).
\ee
As already discussed in Section \ref{sec:tad}, these ensure cancellations of terms from the quartic induced contributions of the form
\be\label{tadloop}
\laa \frac{\delta \Lt}{\delta\R'} \left[ \frac{\delta \Lt}{\delta\R'},\Rh_{\bp}\right]\Rh_{\bp'}\raa  = [\Rh',\Rh_{\bp}]\laa \frac{\delta \Lt}{\delta\R'}\Rh_{\bk} \Rh_{\bp},
\raa
\ee
in which two legs are contracted inside $\frac{\delta \Lt}{\delta\R'}$ on the right-hand side. Those are the only one not needed to build, out of the loop, the connected three-point functions of the form $\partial_{\tau_1}\llangle \Rh_{\bk} \Rh_{\bk'} \Rh_{\bp} \rrangle$. 
By adding Eq. \eqref{tadloop} to the quartic-induced terms, the coefficients highlighted in red in Eqs. \eqref{q1}-\eqref{q2}-\eqref{q3} all become equal to $2$.

\subsection*{One-loop corrections from quartic diff. induced Hamiltonian}
Finally, from the diffs. induced Hamiltonian in Eqs. \eqref{quarticdiffset}-\eqref{quarticdiffset1}, we use the notation: 
\begin{align}
\langle \Rh_{\bp}\Rh_{\bp'}\rangle_{\Ha^{(4)}_{\mathrm{diff},\,X}}\,\equiv\,i\int^{\tau}d\tau_{1}\langle[\Ha^{(4)}_{\mathrm{diff},X}(\tau_1),\hat{\R}_{\bp}\hat{\R}_{\bp'}]\rangle,
\end{align}
and find:
\begin{align}\label{diff1a}
\langle \Rh_{\bp}\Rh_{\bp'}\rangle_{\Ha^{(4)}_{\mathrm{diff},\,A}}\,&=|\R_p|^2  \int^{\tau}d\tau_1\int\frac{d\bk}{(2\pi)^3} \,\eta\cdot \left(   -2 k^2 |\R_k|^2 g_p(\tau,\tau_1)\right)\\[1mm] \label{diff2a}
\langle \Rh_{\bp}\Rh_{\bp'}\rangle_{\Ha^{(4)}_{\mathrm{diff},\,B}}\,&=|\R_p|^2  \int^{\tau}d\tau_1\int\frac{d\bk}{(2\pi)^3} \, \left(   4 \frac{k^2}{a H} |\R_k|^2 \right)\\[1mm] \label{diff3a}
\langle \Rh_{\bp}\Rh_{\bp'}\rangle_{\Ha^{(4)}_{\mathrm{diff},\,C}}\,&=|\R_p|^2  \int^{\tau}d\tau_1\int\frac{d\bk}{(2\pi)^3} \,\left(   -4 \frac{ k^2}{ a H} \Rea(\R_k\R_k'^*)g_p(\tau,\tau_1)\right).
\end{align}
\subsection{One-loop corrections as total derivative terms}

After the length computations leading to the results of the previous two subsections, it would be an almost hopeless task to recognize in their sum a total derivative (with respect to the momentum), unless one travels with a guiding flame.

In particular, from the discussion of Sec. \ref{sec:oneloopthree}, we learned that two legs of the would be three-point function inside the loop originate from operators in the first nested commutator, i.e. $\Ha^{(3)}(\tau_1)$, of the in-in two vertex diagram of Eq. \eqref{twonested}. 
Then, if at least one of these two legs contains an operator with a time derivative, we know from Section \ref{timederandquartic} that the contributions from the quartic-induced Hamiltonian are crucial for providing the additional terms necessary to build three-point functions with time derivative operator inside the loop.

Armed with that, we start combining the results from the cubic interaction: $[c+d+e, c]$ in Eqs. \eqref{cd,c}-\eqref{e,c}\footnote{This leads to the contact in-in formula for the full Hamiltonian considered applied in turns to one external leg $\Rh_{\bp}$ and two $\delta\Lag^{(3)}_c/\delta \R'$.} to the subset of quartic induced Hamiltonian contributions detailed below.
First, to keep separate the contributions induced by the quartic terms explicitly dependent on $\eta$ from the Hamiltonian in Eq. \eqref{secondset} and the one from $\Ham_{\rm e}$, one may rely on the breaking in Eqs. \eqref{quarticinducedset3}-\eqref{quarticinducedset4}. That leads to a splitting for the following contributions: 
\be\label{spe}
\llangle \zeta_{\bp}  \zeta_{\bp'}\rrangle_{H^{(4)}_{3,A}} =- \frac{\eta}{3}\llangle \zeta_{\bp}  \zeta_{\bp'}\rrangle_{H^{(4)}_{3,A}}+ \frac{3+\eta}{3}\llangle \zeta_{\bp}  \zeta_{\bp'}\rrangle_{H^{(4)}_{3,A}} \equiv \llangle \zeta_{\bp}  \zeta_{\bp'}\rrangle_{H^{(4)}_{3,A_1}}+ \llangle \zeta_{\bp}  \zeta_{\bp'}\rrangle_{H^{(4)}_{3,A_2}},
\ee
and
\be\label{spe2}
\llangle \zeta_{\bp}  \zeta_{\bp'}\rrangle_{H^{(4)}_{3,D}} =- \frac{\eta}{6}\llangle \zeta_{\bp}  \zeta_{\bp'}\rrangle_{H^{(4)}_{3,D}}+ \frac{6+\eta}{6}\llangle \zeta_{\bp}  \zeta_{\bp'}\rrangle_{H^{(4)}_{3,D}} \equiv \llangle \zeta_{\bp}  \zeta_{\bp'}\rrangle_{H^{(4)}_{3,D_1}}+ \llangle \zeta_{\bp}  \zeta_{\bp'}\rrangle_{H^{(4)}_{3,D_2}}.
\ee
From each of the two previous equations we take the first term on the right-hand side and sum them to the $[c+d+e,c]$ contributions from the cubic interactions.\footnote{The second term on the right-hand of Eqs. \eqref{spe}-\eqref{spe2} will then be added to the $[c+d+e,e]$ contributions from the cubic interactions.} A bit of algebra leads to:
\begin{align} 
\nonumber
\p_\R^{\rm 1-loop}(p)|_{c}&\equiv \frac{p^3}{2\pi^2}\left(\llangle \zeta_{\bp}\zeta_{\bp'}\rrangle_{[c+d+e,c]}+  \llangle \zeta_{\bp}  \zeta_{\bp'}\rrangle_{H^{(4)}_{3,A_1}}+\llangle \zeta_{\bp}  \zeta_{\bp'}\rrangle_{H^{(4)}_{3,D_1}}\right)\\[1mm] &=  \p_\R^{\rm tree}(p)   \int^{\tau}d\tau_1 \int \frac{d\bk}{(2\pi)^3} \eta\cdot\left( \mathcal{A} -\,g_p(\tau,\tau_1)\,\mathcal{B} \right).
\end{align}
We now turn to the cubic contributions $[c+d+e,d]$ from Eqs. \eqref{cd,d}-\eqref{e,d}. Since $\mathcal{H}^{(3)}_d$ has no time derivative, we do not need any quartic induced contributions to build three-point functions out of this piece. However, we do need a quartic diff. induced contribution to use the consistency relations, this is the one given in Eq. \eqref{diff1a} in $\mathcal{H}^{(3)}_d$. We obtain:
\begin{align}\nonumber
\p_\R^{\rm 1-loop}(p)|_{d}&\equiv \frac{p^3}{2\pi^2}\left( \llangle \zeta_{\bp}\zeta_{\bp'}\rrangle_{[c+d+e,d]}+  \llangle \zeta_{\bp}  \zeta_{\bp'}\rrangle_{H^{(4)}_{\mathrm{diff},A}} \right) \\[1mm]&=  \p_\R^{\rm tree}  (p) \int^{\tau}d\tau_1 \int \frac{d\bk}{(2\pi)^3} \eta\cdot g_p(\tau,\tau_1)\, \mathcal{C},
\end{align}
Finally we consider the $[c+d+e,e]$ cubic contributions. Since the operator $\mathcal{H}^{(3)}_e$ is the more involved once expanded, containing both spatial and time field derivatives, several additional terms have to be included to use the consistency relations. These are precisely given by all left quartic contributions we have not summed up at this point. After tedious rewriting one can combine all left terms and write:

\begin{align}\nonumber
\p_\R^{\rm 1-loop}(p)|_{e}&\equiv \frac{p^3}{2\pi^2}\left(\llangle \zeta_{\bp}\zeta_{\bp'}\rrangle_{[c+d+e,e]}+   \llangle \zeta_{\bp}  \zeta_{\bp'}\rrangle_{H^{(4)}_{3,A_2}}+   \llangle \zeta_{\bp}  \zeta_{\bp'}\rrangle_{H^{(4)}_{3,B}}+\llangle \zeta_{\bp}  \zeta_{\bp'}\rrangle_{H^{(4)}_{3,C}}\right.\\[1mm]\nonumber \quad & \left.+\llangle \zeta_{\bp}  \zeta_{\bp'}\rrangle_{H^{(4)}_{3,D_2}}+\llangle \zeta_{\bp}  \zeta_{\bp'}\rrangle_{H^{(4)}_{3,E}}+\llangle \zeta_{\bp}  \zeta_{\bp'}\rrangle_{H^{(4)}_{\mathrm{diff},B}} + \llangle \zeta_{\bp}  \zeta_{\bp'}\rrangle_{H^{(4)}_{\mathrm{diff},C}}\right) \\[1mm]
=  \p_\R^{\rm tree}(p)  &\int^{\tau}  \int \frac{d\bk}{(2\pi)^3} \left(-(\eta+3)\cdot \mathcal{A}+\frac{3}{a H}\mathcal{B}-\frac{2}{a H}\mathcal{C} + g_p(\tau,\tau_1)\cdot\left( (\eta+3)\mathcal{B} +\frac{1}{a H}\mathcal{D}\right) \right).
\end{align}
By summing all three contributions one arrive at the following expression, summarizing all relevant one-loop contributions to long-wavelength modes from a transient non-slow-roll phase of constant $\eta$:
\be
\p_\R^{\rm 1-loop}(p)=  \p_\R^{\rm tree}  \int^{\tau}  \int \frac{d\bk}{(2\pi)^3} \left(-3\cdot \mathcal{A}+\frac{3}{a H}\mathcal{B}-\frac{2}{a H}\mathcal{C} + g_p(\tau,\tau_1)\cdot\left( 3\mathcal{B} +\eta\mathcal{C}+\frac{1}{a H}\mathcal{D}\right) \right).
\ee
By using the established consistency conditions in Eqs. \eqref{A}-\eqref{B}-\eqref{C}-\eqref{D}, one arrives explicitly at Eq. \eqref{scaleinvariant2}:
\begin{align}
\p_\R^{\rm 1-loop}(p,\tau)&=  \p_\R^{\rm tree} (p,\tau) \int^{\tau}_{\tauf} d\tau_1  \int d k\, C(k,\tau_1),
\end{align}
with $C$ given by the following total derivative:
\begin{align}\nonumber
C(k,\tau_1)&= \frac{d}{d k}\left( 3 \partial_{\tau_1}\p^{\rm tree}_\R(k,\tau_1)  - \frac{3}{a H} \p^{\rm tree}_{\R'}(k,\tau_1) + \frac{2}{a H} k^2 \p^{\rm tree}_\R(k,\tau_1)  \right) \\[1mm]
&+ \frac{d}{d k}\left( g_p(\tau,\tau_1) \left( -3 \p^{\rm tree}_{\R'}(k,\tau_1) -\eta k^2 \p^{\rm tree}_\R(k,\tau_1) -\frac{1}{a H} \partial_{\tau_1}(k^2 \p^{\rm tree}_\R(k,\tau_1) )\right)\right).
\end{align}
Integrating over momenta, we obtain a result that is \textit{independent on the enhanced short scales}, i.e. it depends solely on arbitrary IR and UV momenta far away from the scales of the enhancement. As already mentioned, the latter is the contribution from deep sub-horizon modes whose vanishing is ensured in dimensional regularization. 
The former contribution is suppressed by the ratio of the infrared scale \( k_{\rm IR} \) to the scales associated with the times at which the interactions are considered. This suppression arises because, as expected from the structure of the cubic Lagrangian --see \eqref{main} and recalling $g_p \propto \tau_1$ in the long-wavelength limit-- all terms in parentheses vanish on super-horizon scales.

\section{Conclusions}

The physics of inflation at short scales stands as a \textit{terra incognita}. This stems from the challenges associated with experimentally constraining scales corresponding to wavelengths that exited the horizon during inflation much later than those observed in the CMB.

In this work, we investigate the potential influence of short-scale modes on arbitrary large-scale modes, particularly whether the former could lead to  
large one-loop corrections to the latter. This might challenge the predictability of inflationary scenarios relevant to primordial black hole formation and gravitational wave production, as well as the overall consistency of the inflationary paradigm.

In the context of transient non-slow-roll dynamics, we have explicitly shown that loop corrections to long-wavelengths from short modes are absent. Notably, loop corrections, whose contribution relative to tree level is scale-invariant, \textit{do not} influence the power spectrum of arbitrary large scales. Consequently, the comoving curvature perturbation on scales much larger than the horizon preserves its time-independence at one-loop. 

To achieve our goal, we build upon the reasoning presented in \cite{Pimentel:2012tw} (and in this context, in \cite{Tada:2023rgp}), and derive a series of new results that are essential for drawing our conclusion, which we briefly recap below.

We highlight the importance, in general, of quadratic induced diagrams from the tadpole counterterms--see Sec. \ref{sec:tad}-- to rewrite the one-loop contributions as integral of three-point functions.
After recalling the cubic Lagrangian, we identify the terms in the quartic Hamiltonian which are relevant to compute the one-loop contributions to the long-wavelength power spectrum as induced by non-slow-roll dynamics. In this context, a comparison with results obtained within the framework of the EFT of inflation in \cite{Firouzjahi:2023ahg} (and in \cite{Maity:2023qzw}) would be advisable in the future.
We provide an analytical verification of the Maldacena consistency relations in transient non-slow-roll dynamics.
Specifically for operators with and without time derivatives.
Results are general for arbitrary transient phases of constant $\eta$ (or in general constant spectral index $\nu$) and we highlight the part which depend on the sharpness of the transition, assumed instantaneous here.
Establishing these consistency relations was essential for confirming the structure of the loop corrections.
Extending our previous study in \cite{Fumagalli:2023hpa}, we deliver results, computed with the standard in-in formalism, for the would be leading order one-loop corrections to infrared scales from a transient non-slow-roll
phase of constant $\eta$, by including the effect of the quartic Hamiltonian and additional boundary terms.
Using the verified consistency relations, we show explicitly that all these contributions sum up to a total derivative over the momenta running in the loop.

Our result implies that the dynamics of enhanced short modes is irrelevant for long-wavelength perturbations.
It thus establishes that the foundational predictions of inflation remain intact, even in the face of potential challenges posed by epochs, such as non-slow-roll phases, relevant for PBH and scalar-induced gravitational waves.

Potential avenues for further research include a number of interesting extensions. First, we believe it is crucial, at this stage, a comparison with different approaches entertained in the literature. In particular, studies in the spatially flat gauge \cite{Inomata:2024lud,Ballesteros:2024zdp} and within the EFT of inflation \cite{Firouzjahi:2023ahg}. Also linking to analysis in the separate universe approach \cite{Iacconi:2023ggt} or the large $\eta$ approach advocated in \cite{Tasinato:2023ukp} would be desirable. Second, in light of the present results, it would be valuable to investigate the impact of enhanced short scales on short scales.
This could provide a way to theoretically constrain scenarios relevant for PBHs formation. Third, we are inclined to think that exploring relative scale-invariant corrections in more complex scenarios, such as multifield models or higher-loop corrections, could yield additional insights into the fine structure of the inflationary theory.

Future studies may delve into these aspects. For now, let us wonder about an explicit manifestation of the remarkable interplay between the non-linearities of general relativity, the cubic and quartic Hamiltonians, the consistency relations and the tadpole counterterms. All these elements conspiring together to reveal something that might have seemed intuitive at first glance.\\

\noindent \textbf{Note added.--}While this work was being finalized, similar claims were made in Ref. \cite{Kawaguchi:2024rsv}. Although the authors follow a very different approach (the path integral formalism),
making direct comparison challenging at this stage, our conclusions appear to be in agreement where they overlap.

\section*{Acknowledgments}

The author is grateful to Hassan Firouzjahi, Jaume Garriga, Cristiano Germani, Sadra Jazayeri, Aichen Li, Marco Peloso, Lucas Pinol, Sébastien Renaux-Petel, Flavio Riccardi, Yuichiro Tada, Takahiro Tanaka, Gianmassimo Tasinato and Yuko Urakawa for diverse and useful discussions on the topic of loop corrections in inflation.
 We also thank Jaume Garriga, Sadra Jazayeri, Lucas Pinol, Syksy Rasanen and Sébastien Renaux-Petel for comments on a draft of this paper and to help in deciding some nomenclature. 
The research of J.F. is supported by the
grant PID2022-136224NB-C22, funded by MCIN\allowbreak/\allowbreak AEI\allowbreak/\allowbreak 10.13039\allowbreak/501100011033\allowbreak/\allowbreak FEDER,
UE, and by the grant\allowbreak/ 2021-SGR00872.

\appendix

\section{Formula for quartic induced Hamiltonian from cubic Lagrangian}\label{appquarticinduced}

Here we prove explicitly a well known fact. If we truncate the Lagrangian at cubic order in perturbation theory (in the field $\zeta$ and $\zeta'$) the induced Hamiltonian density $\mathcal{H}$ (we remind our notation $H \equiv \int d^3 x \,\mathcal{H}$) acquires a quartic contribution. This can be conveniently expressed as
\be\label{quartic induced}
\mathcal{H}^{(4)}_{3} = \frac{1}{2 (2 a^2 \epsilon)  }\left(\frac{\delta \mathcal{L}^{(3)} (\R,\R')}{\delta \zeta'}\right)^2.
\ee
To prove Eq. \eqref{quartic induced}, let us start from a Lagrangian density of the form
\be
\mathcal{L}= \mathcal{L}^{(2)}(\R,\R') + \mathcal{L}^{(3)}(\R,\R') + ...
\ee
where $\mathcal{L}^{(2)}$ is given by Eq. \eqref{quadraticR}. The Hamiltonian is obtained through a Legendre transform:
\begin{align}
\Ham(\R,\pp) =& \pp\cdot \R' - \Lag= \pp \cdot \R'(\R,\pp) - \Lag(\R,\R'(\R,\pp))
\end{align}
where the conjugate momentum
\be \label{mom}
 \pp = \frac{\delta \Lag}{\delta \R'}=\frac{\delta \Lag^ {(2)}}{\delta \R'} + \frac{\delta \Lag^{(3)}}{\delta \R'} + ...
\ee
is non-linear in $\R$ and $\R'$ because of $\Lt$.
To find the interaction Hamiltonian up to a given order in perturbation theory in terms of $\R$ and $\R'$ one may go through the following algorithm: a) expand up to a given order in $\R$ and $\pp$ by iteratively inverting the non-linear relations between the conjugate momentum and $\R'$ in Eq. \eqref{mom}; b) substitute to $\R$ and $\pp$ the interaction picture field and momentum, meaning the ones solving the eom coming from the quadratic Hamiltonian $\mathcal{H}_{\rm free}$ c) write everything in terms of the interaction picture $\R,\R'$ by simply using the fact that the interaction picture conjugate momentum is linear in $\R'$. 

Let us start by solving iteratively Eq. \eqref{mom} that we rewrite as:
\be\label{conjmom}
\pp = \gamma \R' + \frac{\delta \Lag^{(3)}}{\delta \R'}, \quad \pp_{\rm lin} = 2\epsilon a^2   \R'\equiv \gamma \R'
\ee
$\pp_{\rm lin}$ is the linear (interaction picture) momentum.

At first order one simply gets
\be
\R' = \frac{\pp}{\gamm}, 
\ee
substituting this into definition of the Hamiltonian and truncating at second order one obtains:

\be
\mathcal{H}^{(2)}\equiv \mathcal{H}_{\rm free} = \frac{P^2}{\gamm} - \Lag^{(2)}\left(\R,\R' = \frac{P}{\gamm} \right) = \frac{P^2}{2\cdot\gamm} + a^2\epsilon  (\partial_i\R)^2
\ee
We now write the full Hamiltonian by using the implicit relation $\R'(\R,\pp)$ as given from Eq. \eqref{conjmom}:

\begin{align}
\Ham &=\pp \cdot \R'(\R,\pp) - \Lag^{(2)}(\R,\R'(\R,\pp)) - \Lag^{(3)}(\R,\R'(\R,\pp)) \\[1mm] 
&=\pp \cdot \left(\frac{\pp}{\gamm} - \frac{1}{\gamm}\frac{\delta \Lag^{(3)}}{\delta \R'}\right) - \frac{\gamm}{2}\left(\frac{\pp}{\gamm} - \frac{1}{\gamm}\frac{\delta \Lag^{(3)}}{\delta \R'}\right)^2 + a^2\epsilon   (\partial_i\R)^2 - \Lag^{(3)}(\R,\R'(\R,\pp)) \\[1mm] \label{partresultH}
&=\mathcal{H}_{\rm free} -\frac{1}{2 \gamm}\left( \frac{\delta \Lag^{(3)}}{\delta \R'}\right)^2 - \Lag^{(3)}(\R,\R'(\R,\pp))
\end{align}
the non-linear part coming from the first term cancel order by order with the cross terms of the squared in the second term. Thus, if we are interested in $\mathcal{H}$ at fourth order, we would need the non-linear relation $\R'(\R,\pp)$ only up to quadratic order, i.e. 
\be
\R' = \frac{\pp}{\gamm} -\frac{1}{\gamm} \frac{\delta \Lt}{\delta \R'}\left(\R,\R' = \frac{P}{\gamm} \right).
\ee

Inserting the previous expression in Eq. \eqref{partresultH}, and expanding $\Lt$, one obtains the interaction picture Hamiltonian
\be\label{intHam}
\mathcal{H_{\rm int}} \equiv \mathcal{H}-\mathcal{H}_{\rm free} =  \mathcal{H}^{(3)} + \mathcal{H}^{(4)}_{3}
\ee
where 
\be
\mathcal{H}^{(3)} = - \mathcal{L}^{(3)},\qquad\mathrm{and}\qquad\mathcal{H}^{(4)}_{3} = \frac{1}{2 (2 a^2 \epsilon) }\left(\frac{\delta \mathcal{L}^{(3)}}{\delta \zeta'}\right)^2.
\ee
both expressed in terms of $\R$ and $\R'$ using the linear relation $P = \gamm \R'$. 

\subsection{Induced quadratic Hamiltonian from tadpole counterterms}

Let us now assume the action is augmented with a tadpole counterterm linear in $\R'$:
\be\label{tad}
\Lag_{\rm tad}^{(1)} = c \,\R',\qquad c =- \laa \frac{\delta\Lt}{\delta\R'}\raa,
\ee
where the braket defying $c$ denotes the Wick contraction of the two-field operators in $\frac{\delta\Lt}{\delta\R'}$.
This would ensure the cancellation of all 1-non-PI diagrams in which two legs from the same cubic interaction are contracted, with the latter having at least one time derivative, i.e. \(\Hamt = -\R' \frac{\delta\Lt}{\delta\R'}\).
The conjugate momentum will now acquire a new contribution $P\supset c$. By using the same steps as in the previous section, one finds the following additions to the interaction Hamiltonian in Eq. \eqref{intHam}:
\be
\mathcal{H}_{\rm int} = -\Lag_{\rm tad}^{(1)} + \Ham^{(2)}_{1}.
\ee
The first term is the expected tadpole counterterm operator from the Lagrangian in Eq. \eqref{tad}, while the second term is a ``quadratic induced Hamiltonian" from the tadpole counterterm and it is given explicitly by:
\be
\Ham^{(2)}_{1} = \frac{c}{2 a^2\epsilon  }\, \frac{\delta\Lt}{\delta\R'}.
\ee

\section{One-loop corrections and boundary terms without time derivatives}\label{APPboundarynoder}
We prove here the following general statement:
one-loop corrections to the power spectrum from cubic boundary interactions without time derivatives cancel exactly with the contributions from diagrams obtained with their correspondent quartic induced Hamiltonian. This was also shown, in a different way, in \cite{Braglia:2024zsl}.

We decompose the Lagrangian density into two parts:
\begin{equation}\label{splittingLp}
\mathcal{L} = \Lagp + \mathcal{L}_{\mathrm{R}},
\end{equation}
where $\Lagp$ represents the generic ``boundary term without time derivatives," which can be expressed schematically as follows:\footnote{In the decomposition discussed in Section \ref{sec:Hamiltonian}, these terms appear on the second line of Equation \eqref{boundary}.}
\begin{equation}
\Lagp = \frac{d}{d\tau} \widetilde{\Lagp}, \qquad \widetilde{\Lagp} = \frac{\lambda}{3}\mathcal{\R}^3,
\end{equation}
and $\lambda$ denotes a time-dependent coupling. We disregard spatial derivatives, as they do not affect the current discussion. The term $\mathcal{L}_{\mathrm{R}}$ encompasses all remaining contributions.
 
The part of the quartic Hamiltonian induced by the presence of $\Lagp$ is (Eq. \eqref{quartic induced}): 

\begin{align}\label{quarticinducedboundary}
\Ham^{(4)}_3 &\equiv \Ham^{(4)}_{3,\,\partial\partial} + \Ham^{(4)}_{3,\,\partial\,\mathrm{R}}
\end{align}
where
\begin{align}\label{quarticpp}
\Ham^{(4)}_{3,\,\partial\partial}&= \frac{1}{2(2a^2 \epsilon )} \left(\frac{\delta\Lagp}{\delta\R'}\right)^2 = \frac{1}{2(2a^2 \epsilon )}\lambda^2\R^4  \\[1mm] \label{quarticpr}
\Ham^{(4)}_{3,\,\partial\,\mathrm{R}} &= \frac{1}{2(2a^2 \epsilon )}\cdot 2\,\frac{\delta\Lagp}{\delta\R'}\frac{\delta\Lag_{\mathrm{R}}}{\delta\R'}  = \frac{1}{2(2a^2 \epsilon )} 2\lambda\,\R^2\frac{\delta\Lag_{\mathrm{R}}}{\delta\R'}.
\end{align}
We prove in the following that a) one-loop diagrams obtained by inserting two cubic total derivative terms cancel with the one-loop contribution from the quartic Hamiltonian \eqref{quarticpp} and b) diagrams obtained by inserting one cubic total derivative term and one cubic generic term cancel with the diagram derived from the quartic Hamiltonian \eqref{quarticpr}.

We start by deriving two useful expressions when dealing with boundary terms within nested commutators in the in-in formula.

\subsection{Useful formula in presence of interaction with boundary terms}
Let us consider two (pieces of the) Lagrangian density, where at least one is given by a total derivative term, i.e.
\be
\Lag_i\equiv \frac{d}{d\tau}\widetilde{\Lag_i},\qquad \Lag_j.
\ee
We write the in-in formula using the notation followed in the main text:
\be
\langle \zeta_{\bp} \zeta_{\bp'}\rangle_{[i,j]}\equiv -\int^{\tau} d\tau_1\int^{\tau_1}d\tau_2 [\Lag_i(\tau_2),[\Lag_j(\tau_1),\zeta_{\bp}\zeta_{\bp'}]],
\ee
where integrals over spatial coordinates are understood.
One can derive the following useful formulas
\be\label{formula1}
\langle \zeta_{\bp} \zeta_{\bp'}\rangle_{[i,j]} + \langle\zeta_{\bp} \zeta_{\bp'}\rangle_{[j,i]}  =\int^{\tau} d\tau_1[\zeta_{\bp}\zeta_{\bp'},[\widetilde{\Lag}_i(\tau_1),\Lag_j(\tau_1)]]
\ee
and (as a special case)
\be\label{formula2}
\langle \zeta_{\bp} \zeta_{\bp'}\rangle_{[i,i]} = \frac{1}{2}\int d\tau_1 [\zeta_{\bp}\zeta_{\bp'},[\widetilde{\Lag}_i(\tau_1),\Lag_i(\tau_1)]].
\ee
\textbf{Proof}
\begin{align}\nonumber
&\langle \zeta_{\bp} \zeta_{\bp'}\rangle_{[i,j]} + \langle\zeta_{\bp} \zeta_{\bp'}\rangle_{[j,i]}   \\[1mm]
 \nonumber
=& -\int^{\tau} d\tau_1 [\widetilde{\Lag}_i(\tau_1),[\Lag_j(\tau_1),\zeta_{\bp}\zeta_{\bp'}]] - \int^{\tau}d\tau_2\int_{\tau_2}^{\tau}d\tau_1 [\Lag_j(\tau_2),[\Lag_i(\tau_1),\zeta_{\bp}\zeta_{\bp'}]]\\
=&- \int^{\tau} d\tau_1 [\widetilde{\Lag_i}(\tau_1),[\Lag_j(\tau_1),\zeta_{\bp}\zeta_{\bp'}]] + \int^{\tau}d\tau_1 [\Lag_j(\tau_1),[\widetilde{\Lag_i}(\tau_1),\zeta_{\bp}\zeta_{\bp'}]].
\label{part}
\end{align}
On the first step we have integrated the total derivative on the first term and made, for the second term, the change of variable
\begin{align}\label{integralred}
\int^{\tau}d\tau_1\int^{\tau_1}d\tau_2 = \int^{\tau}d\tau_2\int_{\tau_2}^{\tau}d\tau_1,
\end{align}
allows us to integrate the total derivative term also in the second term.
From Eq. \eqref{part}, we apply the Jacobi identity to the commutator in the second term and arrive at Eq. \eqref{formula1}. Eq. \eqref{formula2} is a special case where both Lagrangians density are total derivative terms.

\subsection{Boundary - Boundary}
The one-loop contributions obtained by inserting two cubic total derivative terms without time derivatives can be written using Eq. \eqref{formula2} as:
\be
\langle \Rh_{\bp} \Rh_{\bp'} \rangle_{[\partial,\,\partial]} = \frac{1}{2}\int^{\tau} d\tau_1 [\R_{\bp}\R_{\bp'},[\widetilde{\Lagp}(\tau_1),\Lagp(\tau_1)]] 
\ee
In the equal time commutator above, the term from $\Lagp$ containing derivative of the coupling vanishes and we have 
\be
[\widetilde{\Lagp}(\tau_1),\Lagp(\tau_1)] = \frac{\lambda^2}{3}\cdot 3 \Rh^4 [\Rh,\Rh '],
\ee
Using the equal-time commutation relation \be \label{comapp}
[\Rh,\Rh ']'= \frac{i}{2 a^2 \epsilon   },
\ee
one arrives at

\be
\langle \Rh_{\bp} \Rh_{\bp'} \rangle_{[\partial,\,\partial]} =\int^{\tau} d\tau_1 \frac{i}{2(2a^2\epsilon )}\,\lambda^2\cdot[\Rh_{\bp}\Rh_{\bp'},\Rh^4].
\ee
It is immediate to note that the expression above is just equal to minus the contribution from the quartic interaction in Eq. \eqref{quarticpp}, i.e. 
\begin{align}
\langle \Rh_{\bp} \Rh_{\bp'} \rangle_{H^{(4)}_{\partial\,\partial}} &\equiv i \int d\tau_1 [\Ham^{(4)}_{3\,\partial\partial},\Rh_{\bp}\Rh_{\bp'}] = i \int d\tau_1 \left[  \frac{1}{2(2a^2 \epsilon )}\,\left(\frac{\delta\Lagp} {\delta\R'}\right)^2, \Rh_{\bp}\Rh_{\bp'}\right]\\[1mm]
&= \int d\tau_1  \frac{i}{2(2a^2 \epsilon )} \left[ \lambda^2\Rh^4, \Rh_{\bp}\Rh_{\bp'}\right] = - \langle \Rh_{\bp} \Rh_{\bp'} \rangle_{[\partial,\,\partial]} .
\end{align}
\subsection{Boundary - Rest}
The one-loop contributions obtained by inserting one cubic total derivative term without time derivative and one general interaction can be written using Eq. \eqref{formula1} as:
\be
\langle \Rh_{\bp} \Rh_{\bp'} \rangle_{[\partial,\,R]}+ \langle \Rh_{\bp} \Rh_{\bp'} \rangle_{[R,\,\partial]} = \int d\tau_1 \,\frac{\lambda}{3}[\Rh_{\bp}\Rh_{\bp'},[\Rh^3,\Lag_{\rm R}]].
\ee
Similarly as before, one can further re-express the contribution above by using
\be\label{simm}
[\Rh^3,\Lag_{\rm R}] = \Rh [\Rh,\Lag_{\rm R}]\Rh +  [\Rh,\Lag_{\rm R}]\Rh^2 + \Rh^2 [\Rh,\Lag_{\rm R}],
\ee
\be
[\Rh,\Lag_{\rm R}] = [\Rh, \Rh']\frac{\delta\Lag_{\mathrm{R}}}{\delta \R'},
\ee
and the commutation relation in Eq. \eqref{comapp}. At this stage, one realizes that
\begin{align}
\langle \Rh_{\bp} \Rh_{\bp'} \rangle_{H^{(4)}_{\partial\,\mathrm{R}}} &\equiv i \int d\tau_1 [\Ham^{(4)}_{3\,\partial\,\mathrm{R}},\Rh_{\bp}\Rh_{\bp'}] = i \int d\tau_1 \,\left[ 2\cdot \frac{1}{2(2a^2 \epsilon )}\,\frac{\delta\Lagp} {\delta\R'} \left(\frac{\delta\Lag_{\mathrm{R}}} {\delta\R'}\right), \Rh_{\bp}\Rh_{\bp'}\right] \\[1mm]
&= \int d\tau_1  \frac{i}{2(2a^2 \epsilon )} \left[ \lambda^2\Rh^2 \frac{\delta\Lag_{\mathrm{R}}} {\delta\R'}, \Rh_{\bp}\Rh_{\bp'}\right] = - \langle \Rh_{\bp} \Rh_{\bp'} \rangle_{[\partial,\,\mathrm{R}]}-  \langle \Rh_{\bp} \Rh_{\bp'} \rangle_{[\mathrm{R},\, \partial]}.
\end{align}
once operators from the same vertex have been symmetrized in the second-to-last expression. 

\section{Derivatives of background quantities}\label{backquant}
We report here a non-exhaustive list of relations for derivatives of background quantities used (sometimes implicitly) in the main text:
\be
\left(\frac{1}{aH}\right)' = \epsilon - 1,\qquad \left(\frac{1}{aH}\right)''=\epsilon' = \epsilon\eta a H 
\ee
\be
(a^2 \epsilon)' = \epsilon a^2 (a H) (2 +\eta),\quad
(a^2 \epsilon)'' = 2\epsilon a^2 (a H)^2 (3 +2\eta-\epsilon)
\ee
\be
\left(a H (2 +\eta)\right)' = 2(a H)^2(1-\epsilon-\eta^2/2)
\ee
\be
a^2\epsilon \left(\frac{1}{a^2\epsilon a H}\right)'=\epsilon-3-\eta.
\ee

\bibliographystyle{JHEP}
\bibliography{Biblio}
\end{document}